# The development of battery storage systems in Germany: A market review (status 2023)

Jan Figgener[a,b,c,d*], Christopher Hecht[a,b,c], David Haberschusz[a,b,c,d], Jakob Bors[a,b],
Kai Gerd Spreuer[a,b], Kai-Philipp Kairies[d], Peter Stenzel[e], and Dirk Uwe Sauer [a,b,c,d,f]

[a] Institute for Power Electronics and Electrical Drives (ISEA), RWTH Aachen University, Germany

[b] Institute for Power Generation and Storage Systems (PGS), E.ON Energy Research Center (E.ON ERC), RWTH Aachen University, Germany

[c] Juelich Aachen Research Alliance, JARA-Energy, Germany

[d] ACCURE Battery Intelligence GmbH, Germany

[e] TH Köln, Cologne Institute for Renewable Energy (CIRE), Germany

[f] Helmholtz Institute Münster (HI MS), IEK-12, Forschungszentrum Jülich, Germany

*Corresponding author at Institute for Power Electronics and Electrical Drives (ISEA)

Mathieustraße 10, 52074 Aachen, Germany | Mail: jan.figgener@rwth-aachen.de

***Abstract*** **- The market for battery storage systems (BSS) has been growing rapidly for years and will multiply in the future. This fast growth leads to a lack of information regarding current developments. With this extension of our previous works, we contribute key figures for model parametrization and political decision-making and depict the market development in Germany, one of the leading storage markets worldwide. In empirical analyses, we evaluate and combine all major public databases on national stationary and mobile storage as well as our databases from subsidy programs and extend the insights by literature research and bilateral industry exchange. In comparison to 2021, the market for home storage systems (HSS) grew by 52% in terms of battery energy in 2022 and is by far the largest stationary storage market in Germany. We estimate that about 220,000 HSS (1.9 GWh / 1.2 GW) were installed solely in 2022. The emerging market for industrial storage systems (ISS) grew by 24% in 2022, with a total of 1,200 ISS (0.08 GWh / 0.04 GW) installed. The market for large-scale storage systems (LSS) increased strongly by 910% with 47 LSS (0.47 GWh / 0.43 GW) commissioned. The electric vehicle (EV) market grew with 693,000 new EV (27 GWh / 43 GW (DC) / 4.5 GW (AC)) by 34% in terms of battery energy. The number of EV per charging point grew from 9 in 2017 to 23 in 2022. System BSS prices increased significantly in 2022 and were estimated at 1,200 €/kWh for HSS. LSS prices ranged on average from 310 €/kWh to 465 €/kWh. In comparison, if the 2022 BEV prices for the whole vehicle are simply divided by their battery energy, the mean specific BEV system prices range from 800 €/kWh for medium to 1,240 €/kWh for luxury cars. In total, we estimate that over 650,000 stationary BSS with a battery energy of 7.0 GWh with an inverter power of 4.3 GW and 1,878,000 EV with a battery energy of 65 GWh and a DC charging power of 91 GW (12 GW AC) were operated in Germany by the end of 2022. The cumulative battery energy of about 72 GWh is therefore nearly double the 39 GWh of nationally installed pumped hydro storage demonstrating the enormous flexibility potential of battery storage for the energy system.**

## I. INTRODUCTION

This paper is an update of our existing peer-reviewed works [1–4] and extends large parts of the previous analyses.

In current forecasts on the development of the global battery market, everyone agrees: it is going steeply upwards. Nevertheless, the estimates differ significantly from each other and change over time, which is due to different future scenarios, changing regulatory, geopolitical circumstances, and a lack of transparently accessible information as a study on the European storage market explicitly points out [5]. To understand the market dynamics, we can have a look at the past: A 2017 study predicted the cumulative stationary battery world installations for the year 2030 to range between around 100 GWh and 420 GWh depending on the scenario [6]. Another 2017 study estimated cumulative installations of 305 GWh for 2030 [7]. Only a few years later, in 2022, they updated their estimate by a factor of four to 1,194 GWh as the storage world is in flux [8]. According to their estimate, the countries with the most battery installations will be the United States, China, Japan, India, Germany, United Kingdom, Australia, and South Korea among others [8]. In order to contribute to transparency, we evaluate the current developments for Germany while many key figures like prices, use cases, and system design are also valid for larger parts of the world.

*Table 1: Abbreviations sorted alphabetically.*

| Abbreviation | Description |
|---|---|
| aFRR | Automatic frequency restoration reserve |
| ADAC | General German Automobile Club |
| BEV | Battery electric vehicle |
| BSS | Battery storage system |
| CP | Charging point (for electric vehicles) |
| CS | Charging station (for electric vehicles) |
| DB | Database |
| EPR | Energy-to-power (ratio) |
| EV | Electric vehicle |
| FCP | Fast charging point (for electric vehicles) |
| FCR | Frequency containment reserve |
| FMTA | (German) Federal Motor Transport Authority |
| FNA | (German) Federal Network Agency |
| HSS | Home storage system |
| ISEA | Institute for Power Electronics and Electrical Drives |
| ISS | Industrial storage systems |
| LSS | Large-scale storage system |
| PHEV | Plug-in hybrid electric vehicle |
| PV | Photovoltaic |
| SOC | State-of-charge |
| TSO | Transmission system operator |

There are already many publications on the market development of battery storage in the literature. According to our classification in [1], these can be divided into the three areas of (1) institutional publications, (2) peer-reviewed publications, and (3) consultant publications.

(1) The institutional publications are often from renowned organizations such as IRENA [6], the European Commission [5], or the IEA [9]. The reports, however, cover larger geographical areas such as the world or Europe and have either no or only isolated meta-information on individual countries. For the stationary BSS sector, there are two reports specifically for Germany from BVES and BSW-Solar, the two major associations in the storage sector. The BVES report [10] is freely available and provides an overview using parts of our data for the storage market update. The report offers information on a high level and does not provide detailed information on technology development such as battery chemistry. The BSW-Solar publishes single key information, such as installation numbers, in factsheets [11]. For the EV sector, in addition to the global analyses of the IEA [9], there are also reports from German institutions like the Federal Motor Transport Authority (FMTA) [12] or the NOW GmbH [13], which evaluate new registrations of EV in Germany but do not give information on data such as battery energy and charging power.

(2) The peer-reviewed publications are mostly based on the DOE database [14]. This is a recommended database with mostly large BSS projects from all over the world, but unfortunately incomplete for individual countries due to its global focus. In the publications, analyses of the database are sometimes supplemented by further information such as manual research or interviews [15–18]. In addition to a worldwide focus, there are also analyses for individual countries [19, 20], but we are unaware of any detailed analyses for Germany other than our own [1–4]. Another database for BSS around the world is provided by Lorenz Gruber [21] but is not supplemented by any publication.

(3) Many consultant publications are fee-based and therefore not publicly available [22–25]. In some cases, there are also free reports or excerpts [7, 26–29], but the methodology is usually not disclosed and is therefore non-transparent from a scientific point of view.

Motivated by these developments and the adoption of our previous analyses [1–4] in scientific literature, public reports [5, 6], and the press [30, 31], this paper provides an update on the market development of BSS in Germany. In addition to the previously covered markets of home storage systems (HSS), industrial storage systems (ISS), large-scale storage systems (LSS), and EV, we address several new application areas and expand our analyses (see Table 1 for abbreviations and Table 2 for BSS classification).

These expansions include (1) transparent quantification of non-reporting quotas of HSS and LSS to the Federal Network Agency, (2) application areas of both ISS and LSS, (3) economic sectors that operate ISS, (4) results of the first five rounds of the so-called innovation auctions for BSS, (5) detailed price analysis of both BSS and EV, (5) specific EV key metrics such as battery energy, DC charging power, range, and consumption for different vehicle classes, (6) public charging stations, and (7) announced battery production capacities in Germany.

In order to keep selected figures on the stationary storage market up to date, we developed the website "Battery Charts" (www.battery-charts.de) [32] which refers to this paper and accompanies it as supplementary material. Selected EV figures are available analogous on the website "Mobility Charts" (www.mobility-charts.de) [33] as presented in [3].

On the one hand, this paper serves as a transparent primary source to parameterize simulation models. On the other hand, it provides policymakers and industry with information that supports decision-making in a dynamic market environment. Technological and regulatory focus as well as purely data-based and transparent presentations are important characteristics of our study.

## II. Methodology

In this chapter, we describe the methodology. After a description of the different databases, we explain how we arrive at estimates of both the stationary and the mobile storage market.

### II.A. Databases

Our analyses are based on the linkage of several databases. These databases are divided into public and private databases and summarized in Figure 1 and Table 3 with a detailed description.

### II.B. Estimation of the stationary storage market

The Federal Network Agency (FNA) launched its MASTR DB in 2019, where all BSS need to be registered. After its launch, it took a while for storage operators to be aware of the obligatory requirement of BSS registration, which led to many BSS that were not registered [1, 2] (see section III.A for further information). Due to the increasing completeness of the registrations in MASTR DB, the analyses on the stationary battery storage market are based primarily on its evaluation [34]. The analyses are validated and supplemented using the private MONITORING DB, LSS DB [35], and data from bilateral exchange with industry partners. The simplified classification of storage systems is made according to Table 2 following our previous classification in [2]. Further information can be found in our previous works [1, 2, 4].

*Table 2: Storage classification and filters applied to MASTR DB [34].*

| Market | Filter |
|---|---|
| HSS | battery energy ≤ 30 kWh |
| ISS | 30 kWh < battery energy < 1,000 kWh |
| LSS | battery energy ≥ 1,000 kWh; operated by legal entities; voltage level "low voltage (230 V P-N)" only allowed if grid operator has approved entry |
| HSS, ISS, LSS | BSS must be in operation; both energy and power must be registered; 0.1 h ≤ energy-to-power ratio (EPR) ≤ 15 h |

## II.C. Estimation of the mobile storage market

The EV market is analyzed by matching manufacturer key number and type key number of the vehicle stock in the dataset "FZ6" provided by the German Federal Motor Transport Authority (FMTA) [12] with technical vehicle data from ADAC DB [37]. Using this method, over 90% of EV could be identified. For 2022, the "FZ6" dataset has not been published yet, so we used the new registrations (dataset "FZ10" [39]) and scaled them with the published stock number [40].

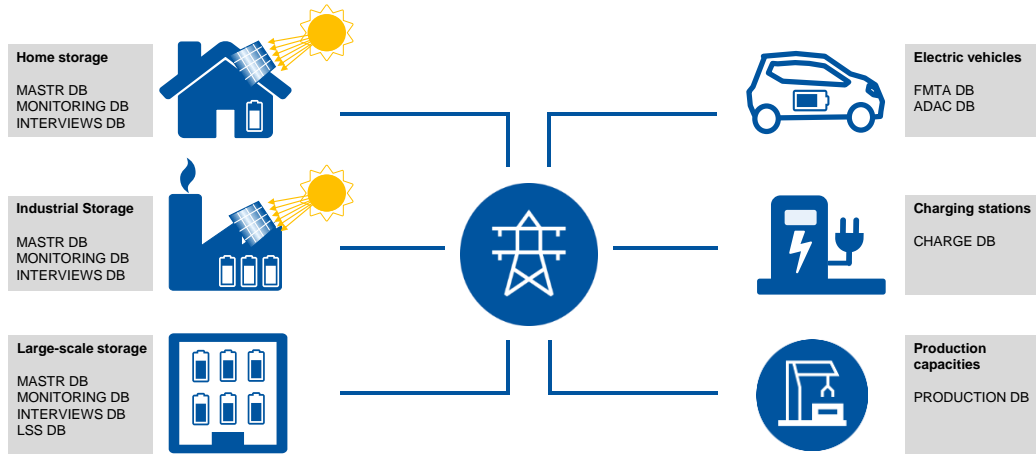

*Figure 1. Overview of the databases used for the analyzed markets. See database descriptions and links in Table 3.*

*Table 3: Description of the databases used.*

| Database | Markets | Description |
|---|---|---|
| *MASTR DB [34] (public)* | *HSS, ISS, LSS* | *The MASTR DB [34] is the public database of the German Federal Network Agency (FNA, German: "Bundesnetzagentur"), where all stationary BSS have to be reported since January 2019. After some initial difficulties, the reported BSS are now almost complete and the database offers a good overview of the market development, especially for newly registered BSS (see section III.A). The data is based on manual entries by private persons, which is why there are regularly confusions of technical quantities and incorrect information. To account for these, consistency checks are performed by the FNA and distribution system operators. In addition, we apply the filters mentioned in Table 2 for a consistent dataset. The collected data captures information about the location, the operator, and the inverter power and energy of the batteries.* |
| *MONITORING DB [1, 2, 4] (private)* | *HSS, ISS* | *Until the beginning of MASTR DB, ISEA had the most comprehensive databases with around 30,000 HSS and ISS through the monitoring [1, 2, 4] of the national subsidy program of the German government (2013-2018) and the monitoring [36] of the local subsidy program of the federal state of Baden-Württemberg (2018-2022). The databases were collected via online questionnaires on subsidized HSS and ISS and are used for more detailed evaluations of the BSS such as technical characteristics of the BSS or additional information such as price development.* |
| *LSS DB [35] (public)* | *LSS* | *The "Forschungszentrum Jülich" compiled the LSS DB before the introduction of MASTR DB by manual research. This data was compared, extended, and aligned with the MASTR DB for this paper.* |
| *INTERVIEWS DB (private)* | *HSS, ISS, LSS* | *Over the years, intensive connections have been established with associations, manufacturers, market research institutes, installers, wholesalers, and order brokers. The regularly conducted bilateral exchanges are used to gather information about the storage market and current developments.* |
| *FMTA DB [12] (public)* | *EV* | *The German Federal Motor Transport Authority (FMTA, German: "Kraftfahrt-Bundesamt") regularly publishes new vehicle registrations in Germany. The BEV and PHEV categories are also listed and the vehicle type codes are given.* |
| *ADAC DB [37] (public)* | *EV* | *The General German Automobile Club (ADAC, German: "Allgemeiner Deutscher Automobil-Club") maintains a database of technical details such as battery inverter power and energy of almost all vehicle models available on the market. Technical characteristics are exposed on the ADAC website and were compiled from there.* |
| *CHARGE DB [38] (public)* | *CS* | *The FNA maintains a register of all public charging stations reported in Germany, including installed charging power and location.* |
| *PRODUCTION DB (private)* | *BSS, EV* | *For the production capacities, we did research of press releases and compare our results to other estimates.* |

This is necessary to account for de-registrations. The found battery energy and charging power were then scaled up to represent 100% of battery electric vehicles (BEV) and plug-in hybrid electric vehicles (PHEV). The evaluations of the public charging infrastructure are carried out exclusively by the statistical evaluation of the CHARGE DB [38] provided by the FNA [38]. For further information on matching the registered EV at FMTA DB [12] to technical data from ADAC DB [37], we refer to Hecht et al. [3], in which we focus on the methodology.

## III. RESULTS AND DISCUSSION

This section depicts the current market development of stationary battery storage, electric vehicles, charging infrastructure, and battery production capacities in Germany.

### III.A. Home storage market in Germany

The home storage system (HSS) market is the largest stationary storage market in Germany and has seen rapid growth in recent years. Figure 2 shows the estimate of annual HSS installations according to battery technologies used. According to our analysis, about 220,000 HSS were installed in 2022, representing market growth of 52% with respect to the previous year. In total, we estimate a stock of 650,000 HSS installations in Germany by the end of 2022.

This estimate is based on analyzing the MASTR DB, where both PV systems and BSS are registered as separate datasets. However, the PV registrations also provide information on if they were installed together with an HSS and we consider them to be more accurate than the HSS registrations themselves. The reason for this is that the feed-in tariff is only paid if the PV system is registered, meaning that virtually all PV systems are included. In the MASTR DB, around 285,000 PV systems with a power between 2 kW and 30 kW were installed in 2022 (status February 2023). From the new residential PV registrations, nearly 75% state to be installed together with an HSS, leading to 210,000 HSS installations with new PV systems. When matching the PV with the HSS registrations of 2022, 95% of the HSS dataset are installed together with a new PV system of 2022 and 5% are retrofits to older PV systems. Taking these retrofits into account, the total HSS estimation reaches about 220,000 HSS for 2022.

The non-reporting quota of HSS is therefore estimated to be around 9% as 202,000 HSS have been registered at the FNA for 2022 by February 2023. However, further analysis shows that this is about the same share of HSS which is typically registered belated after the end of a year. Thus, the MASTR DB is estimated to be the most reliable source to track the market development of stationary battery storage in Germany and can be used from now on without further analyses.

Our estimate is validated through bilateral exchanged data from the DAA [41], which places tens of thousands of PV and HSS contracts annually. In 2022, 66% of the orders included a PV system plus storage, 10% included a PV system explicitly without storage, and 24% of the inquiries had not yet been decided. The retrofits analyzed by DAA also account for 5% to 7% and it is likely that half of the undecided people opt for an HSS, which supports our analysis of 75% of PV systems being installed together with an HSS. Our estimates are in line with other analyses ranging from 197,000 HSS (HTW Berlin [42]), 214,000 HSS (BSW [43]), to 220,000 HSS (EUPD [44]).

The installed HSS are almost completely (98%) equipped with lithium-ion batteries continuing the trend of previous years. While, at the market's beginning, there were high shares of lead-acid batteries, they have nearly disappeared. Other technologies such as redox-flow and salt-water also exist, but are a niche. All non-lithium-ion technologies combined achieve a market share of only 2% in 2022 and within this share, there are also many lithium-ion HSS that were falsely classified as "Others" as manual checks reveal.

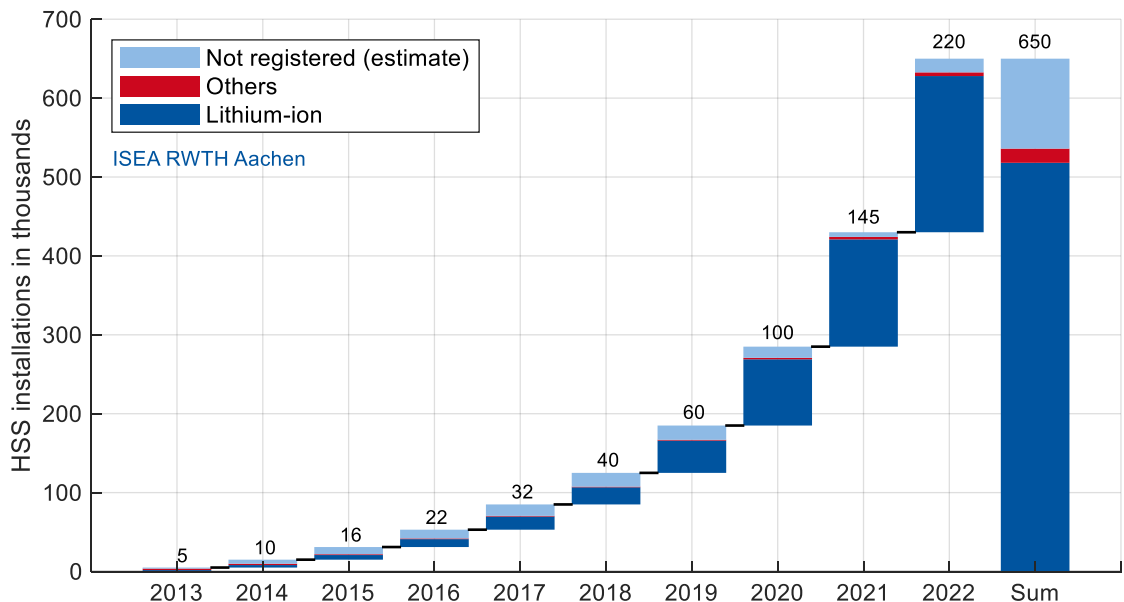

*Figure 2. Estimated number of HSS installations in Germany based on own analyses of MASTR DB [34], and bilateral exchange with installers, retailers, and manufacturers.*

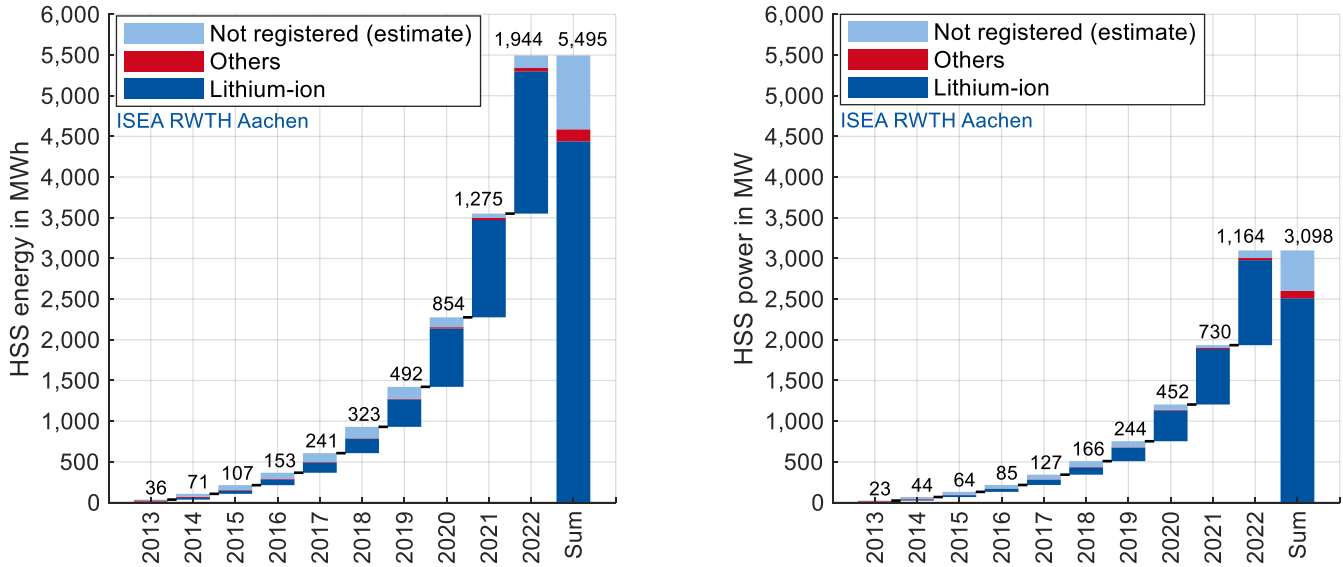

*Figure 3. Estimated HSS battery energy (left) and inverter power (right) in Germany based on own analyses of MASTR DB [34], and bilateral exchange with installers, retailers and manufacturers.*

In 2022, the newly installed battery energy was around 1,944 MWh and the battery inverter power about 1,164 MW (see Figure 3), representing a growth of 52% (energy) and 60% (power), respectively, relative to the previous year. Thus, by the end of 2022, a total HSS battery energy of 5,495 MWh and battery inverter power of 3,098 MW were installed in Germany.

The average energy per system stayed nearly constant at 8.8 kWh in relation to the previous year while the average power increased further from 5 kW to 5.3 kW (see appendix, Figure 18). Especially the segment of HSS above 10 kWh accounts for an increasing share of HSS installations with around 37% of systems, while the majority of HSS is still between 5 kWh and 10 kWh with around 56% of systems (see appendix, Figure 19). Power ratings grew quicker than energy ratings following the preferences of customers to achieve higher self-sufficiency including EV charging, and heat pump operation. The energy-to-power ratio (EPR) for most systems is around 2 h (see appendix, Figure 20).

### III.B. Industrial storage market in Germany

The industrial storage system (ISS) market is the smallest stationary storage market in Germany, but growing.

The year 2022 showed the highest additions recorded until now (see Figure 4). The approximately 1,175 newly registered ISS have an energy of about 84 MWh and a power of 43 MW (see Figure 5). In terms of energy, this is a growth of about 24% over 2021. By the end of 2022, around 3,900 ISS had a cumulative battery energy of 269 MWh and an inverter power of 144 MW.

Due to lack of information about the previous years of the ISS market described in [2] and multiple use cases, we stick solely to the MASTR DB registrations even though there could also be ISS that have not been registered as the BVES suspects [45]. However, in case a similar quote of non-reported ISS is assumed according to the HSS or LSS market (below 10% each), the market is still quite small.

In terms of battery technology, lithium-ion systems continue to lead with 95% of the ISS energy and lead-acid systems account for about 3%. The ISS designated as "Other" in the MASTR DB are also predominantly lithium-ion systems according to manual triage and redox-flow systems account to only less than 1% of the installed energy in 2022.

Overall, lithium-ion batteries have a higher efficiency and longer lifetime in many applications. Nevertheless, lead-acid batteries are particularly suitable for uninterruptible power supply. These systems are partly not included in the dataset as they only have to be registered if they are not exclusively used for uninterruptible power supply [46]. The reason for lead-acid batteries being also used for uninterruptible power supply is that the batteries are on standby most of the time at high states-of-charge (SOC). At high SOC, however, lithium-ion batteries exhibit accelerated aging [47]. This is not the case with lead-acid batteries.

Redox-flow batteries, on the other hand, offer the charm of largely independent dimensioning of energy and power. While energy is determined by tank size, the power is influenced by pumping speed and membrane area [6]. This can also be seen with an energy-to-power ratio (EPR) of 4.3 h for four commercial redox-flow ISS (130 kWh / 30 kW) registered in the MASTR DB and the largest registered redox-flow ISS of 400 kWh and 100 kW. The designs thus correspond to discharge durations twice as long as the average for the overall market. Due to the lower efficiency and the additional susceptibility to error in the form of the pumping system as well as increased maintenance, these batteries currently have only small market shares and they do not achieve any price advantages according to our MONITORING DB.

Most ISS in operation are in the 30 kWh to 100 kWh class (see appendix, Figure 19). This segment accounts for about 86% of the systems. The share of ISS between 100 kWh and 200 kWh accounts for nearly 8% of the systems. ISS above 200 kWh, however, remain the exception among large industrial operations, accounting for about 6% of installed systems. While the median EPR is around 3 h, the energy-weighted mean is below 2 h (see appendix, Figure 20). Thus, especially larger ISS tend to have lower EPR.

We depict the ISS use cases (see appendix, Figure 21) and the sectors (see appendix, Figure 22) this year for the first time. In total, we could assign 74% of the ISS energy to a use case and 68% to a sector. About half of the ISS energy is used for renewable energy (RE) integration (match of MASTR DB for ISS and PV), 10% for a combination of RE integration and EV charging, and 1% solely for EV charging without a PV system (match of MASTR DB for ISS, PV and CHARGE DB). Further, 12% state in the MASTR DB to serve as an uninterruptible power supply.

In terms of the sector, services and public administration is leading with 18% of the ISS energy followed by 10% installed in manufacturing and 8% in the energy supply sector. Trade as well as agriculture, forestry and fishing account for 5% and 4%, respectively. The sectors were chosen according to the NACE classification of the economic activities in the European Community [48].

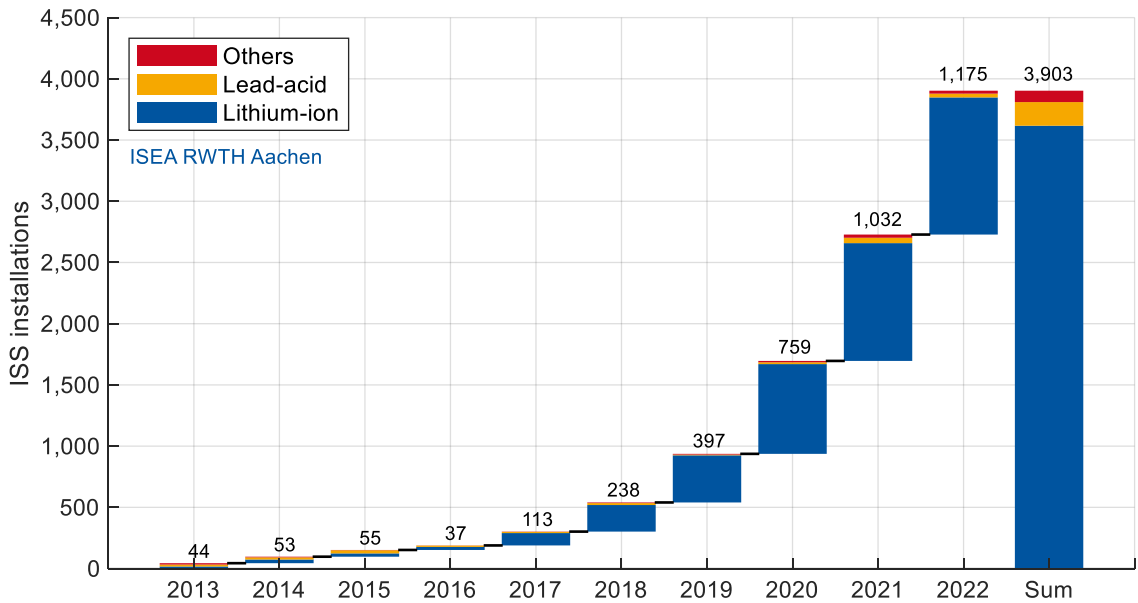

*Figure 4. Estimated number of ISS installations in Germany based on own analyses of MASTR DB [34].*

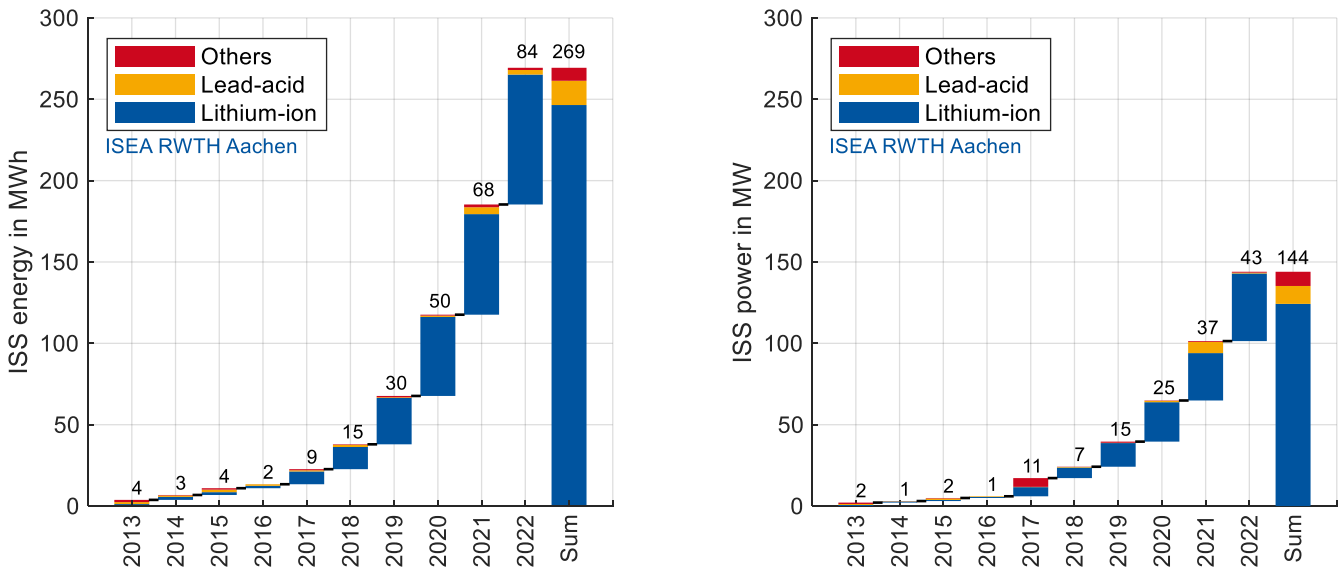

*Figure 5. Estimated ISS battery energy (left) and inverter power (right) in Germany based on own analyses of MASTR DB [34].*

### III.C. Large-scale storage market in Germany

The large-scale storage market (LSS) is the second largest stationary BSS market in Germany. Section III.C.1 discusses the market development while section III.C.2 presents a deep dive into the results of the innovation auctions to give a better understanding of the emerging use case of RE integration. Section III.C.1 includes all LSS from section III.C.2.

#### III.C.1. Market development

After some years of decline, 2022 showed a record of 47 new LSS installations (see Figure 6) with 467 MWh and 434 MW, which corresponds to a growth of 910% in terms of energy additions (see Figure 7). In December 2022 alone, 250 MWh were registered. One reason for this end-of-year rally could be the last chance to register the LSS for the avoidance of grid tariffs according to § 18 StromNEV [49, 50]. At the end of 2022, there were in total 149 LSS with an energy of 1,204 MWh and a power of 1,072 MW installed. For the next two to three years, the public announcements already exceed the whole cumulative LSS installations until the end of 2022 with 1,414 MWh / 1,123 MW.

We estimate that 9% of LSS have not been registered until the end of 2022 when comparing the MASTR DB data with LSS DB and further literature research of press releases and newspapers. The missing registrations for LSS in operation are exclusively LSS that were commissioned before the start of MASTR DB in 2019. In regard to planned LSS projects, we found only six additional projects, although these are quite large (see Figure 7).

Most LSS are below 10 MWh (77%), 18% are between 10 MWh and 20 MWh, and single projects reach up to 80 MWh, while there are projects over 200 MWh announced for the future. After some technologically versatile installations from 2016 to 2019, lithium-ion LSS have now been installed almost exclusively for the third year in a row. With lithium-ion batteries manifesting their dominance in the LSS sector, all stationary and mobile battery markets are overwhelmingly served by the technology. Although some planned projects still state to have "Others" as battery technology or have not yet specified the technology, it is most likely that these projects will be realized as lithium-ion batteries as well.

While the technology does not show any shift, the use cases do as our manual research indicates (see appendix, Figure 23). Besides (1) ancillary services, (2) RE integration, (3) industrial energy supply, (4) multi-use operation including arbitrage trading, (5) grid booster projects, and (6) others gain traction. Figure 23 only shows the use case we considered to be dominating although some are also in multi-use operation. The EPR depends strongly on the use case with about 1 h for ancillary services, 1 h to 2 h for RE integration, 1 h for grid booster and a broad range from 0.5 h to 3 h for industrial energy supply (see appendix, Figure 20).

(1) Ancillary services (in operation: 750 MWh / 685 MW) are still leading. Until 2019, LSS have been built almost exclusively for the provision of frequency containment reserve (FCR) [1, 2], which we evaluate in [51–55]. During this time, however, the prices for FCR initially dropped significantly, primarily due to the increasing saturation of the market volume by battery storage, which is why economic operation became more difficult and the market declined until 2021 [1, 2], but increased in 2022 again [56]. As of January 2023, the German FCR market is 570 MW [57], while already 630 MW of LSS are officially prequalified to provide FCR [57] and national LSS thus need to make use of the option to export 30% of the FCR power to other countries. Although FCR prices are still attractive in early 2023 [58], analogous to all other energy markets, this price level will probably not be sustained in the long term due to high competition and further large projects announces for this market. This is why LSS now also evaluate other possibilities like automatic frequency restoration reserve (aFRR), where the market is three to four times as large as the FCR market and until now only 60 MW of LSS are prequalified [57]. The ancillary service aFRR can be offered either as positive or negative power. LSS can market both directions simultaneously as negative and positive power are never activated at the same time.

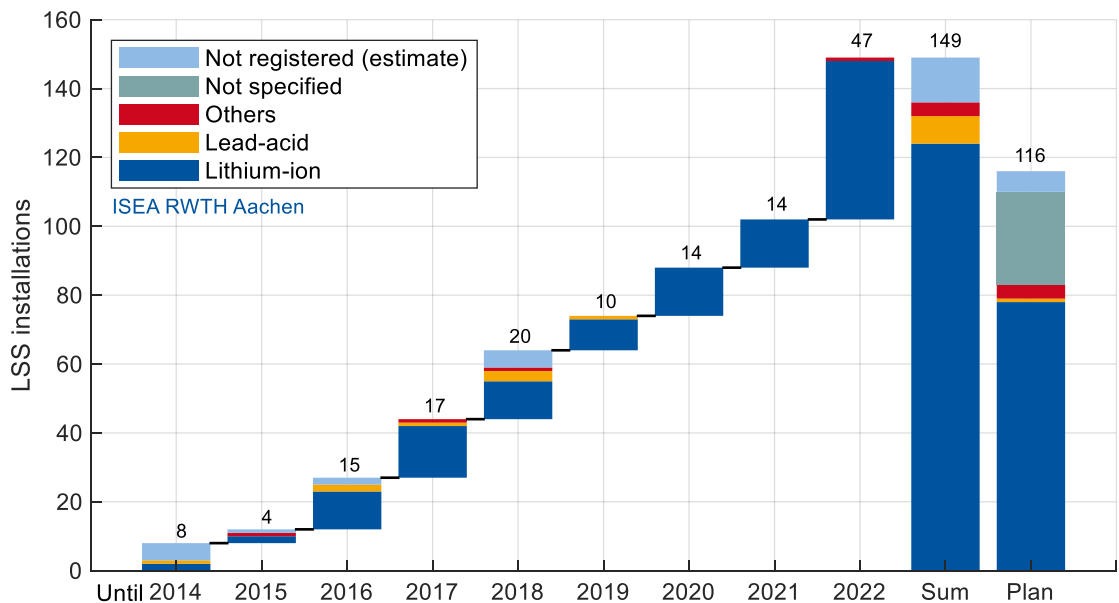

*Figure 6. Estimated number of LSS installations in Germany based on own analyses of MASTR DB [34], LSS DB [35], and further literature research. "Plan" includes all announced LSS. "Not registered" refers to LSS that are not yet registered in MASTR DB.*

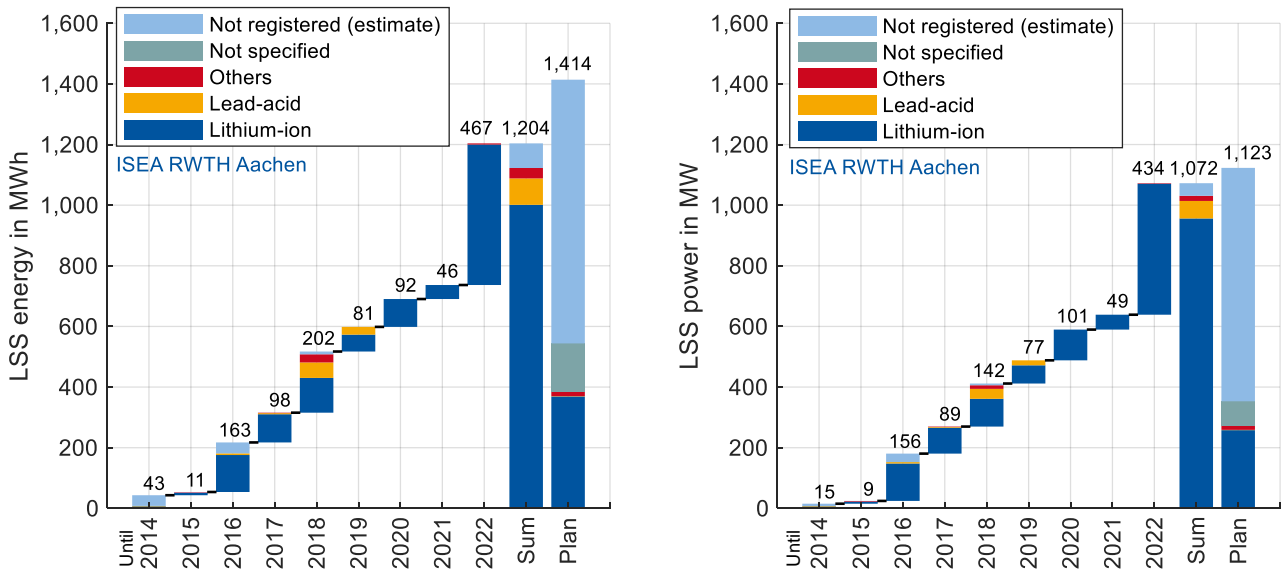

*Figure 7. Estimated LSS battery energy (left) and inverter power (right) in Germany based on own analyses of MASTR DB [34], LSS DB [35], and further literature research. "Plan" includes all announced LSS for the next 2-3 years. "Not registered" refers to LSS that are not yet registered in MASTR DB [34].*

Instead of a single power price of the FCR market, the aFRR market has a power and an energy price, making bidding strategies more complex. The maximal power needs to be provided at least for one hour, so the LSS need to have larger energy-to-power ratios than for FCR provision. The aFRR prices have been increasing over the last years, possibly leading to an attractive alternative [58].

(2) RE integration (in operation: 250 MWh / 200 MW) gained traction in the LSS market and there is already 700 MWh announced for the next years. Most projects are PV and LSS projects and there is only one wind plus LSS project until now within the innovation auction.

(3) Industrial energy supply (in operation: 50 MWh / 50 MW) is on the rise as the first large industrial sites operate LSS. For example, an engine manufacturer operates an LSS with an EPR of 4 h presumably for its self-consumption while another uses a high-power LSS with an EPR of only 30 min most likely for peak shaving. Further, there are LSS that are installed next to power plants to optimize operation.

(4) In multi-use operation (in operation: 150 MWh / 130 MW), LSS switch from one market to another and can combine front-of-meter use cases like ancillary services with behind-the-meter use cases such as self-consumption or peak shaving. The operating company of a couple of LSS advertises on its website multi-use operation of energy arbitrage, load management, peak shaving, voltage stability, and FCR. Especially arbitrage trading is finally happening. While some years ago, this was economically not promising, the high price spreads on the spot market lead to LSS shifting energy from times with low to times with high prices. However, trading strategies have to keep in mind that the energy throughput can be quite high and it is easily possible to exceed the cycle life of a battery. Typically, one to two cycles per day are realistic to not exceed most warranty conditions. As a high cycle depth leads to accelerated aging [51], trading strategies should try to do many small cycles instead of a few large ones.

(5) Grid boosters (planned: 450 MWh / 450 MW) to temporarily relieve grid bottlenecks and save preventive redispatch have been increasingly discussed in recent years. These grid boosters will be among the largest storage projects in the world to date, with several hundred megawatts and megawatt-hours when completed. The planned pilot projects for concept validation by the transmission system operator TenneT, each with coordinated 100 MWh and 100 MW at the Audorf/Süd and Ottenhofen sites, and by the transmission system operator TransnetBW with 250 MWh and 250 MW at the Kupferzell site, were designed by the transmission system operators in the 2030 grid development plan and confirmed by the FNA in 2019 [59]. According to the 2035 network development plan, the two projects P365 and P430 are each in the preparation of the planning and approval process. The expected commissioning of TenneT's project (P365) is planned for 2023 and that of TransnetBW's project (P430) for 2025. In contrast to the LSS for the provision of FCR and the LSS within the scope of the innovation auctions, grid boosters are used exclusively for grid operation management [59].

(6) Others (8 MWh / 6 MW) include two EV charging, one black start, and five LSS that are not specified yet.

Similar to the HSS, the MASTR DB is considered to be a very reliable source for LSS deployments in Germany, especially for systems in operation. Only some LSS that were mostly commissioned before the start of the MASTR DB seem to have forgotten to register their systems retrospectively. However, in terms of announcements one either has to wait for the registration or collect press releases.

### III.C.2. Innovation auctions

The Federal Network Agency (FNA) conducted the first five rounds of the so-called "innovation auctions" from 2020 to 2022 for the RE integration of large solar and wind parks that can bid in combination with a BSS [60]. The basis for the auctions is regulated in the Renewable Energy Sources Act [61] (Section 39j, EEG 2014, with amendments of 21.12.2020 and Section 39n, EEG 2021, respectively) and in the ordinance on innovation auctions [62]. The innovation auction BSS are mostly above 1 MWh, which is why they are listed in the LSS section, although there are also some BSS that would be classified as ISS according to Table 2. As of February 2023, 21 BSS are already in operation, while 84 BSS are in construction according to MASTR DB [2].

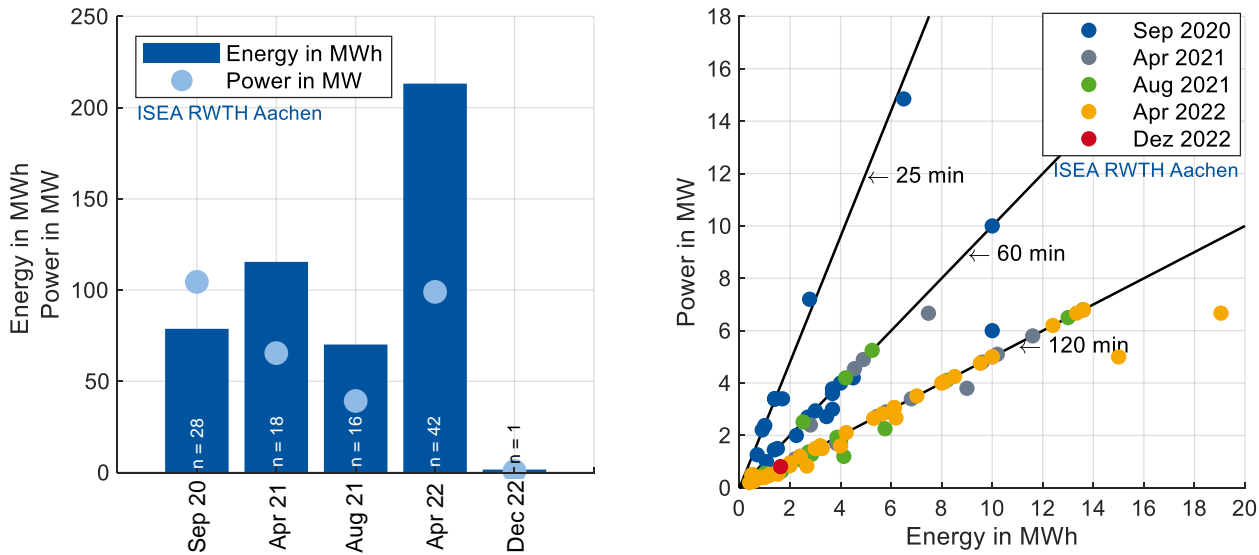

*Figure 8. BSS battery energy and inverter power (left) of the first five innovation auction rounds [60] and the system design (right) of the single BSS. As of February 2023, 21 BSS are already in operation, while 84 BSS are under construction. All BSS are either lithium-ion batteries or not specified.*

The total reported energy of the 105 BSS amounts to 479 MWh while the power adds up to approximately 309 MW (see Figure 8). The largest energy of a single project is 19 MWh with an EPR ratio of nearly 3 h and the largest power (in another project) is 14.9 MW with an EPR ratio of 25 min. Of the 105 BSS, 69 were reported to use a lithium-ion technology and the others are not specified by now. It should be noted that the values shown are preliminary and may change in the course of BSS commissioning.

The BSS power must at least be one third of the RE power. However, the single innovation rounds have differences in the EPR ranging from an average of 0.75 h in September 2020 to an average of 2.15 h in April 2021. The reason for this are changed regulatory requirements after the first round in September 2020. In this first round, it was still debatable whether the BSS should have been designed for one hour or 25 min. The design for one hour is based on the requirements of the transmission system operators for limited energy storage [63]. However, as the BSS do not have to reach the minimum size to participate in the aFRR market as specified in the innovation auctions regulation, some projects are also designed only to meet the test protocol for aFRR. The protocol is used to verify two test calls of ten min of full power each, which requires about 25 min of full power including buffers [63]. These two options clearly show in the project-specific allocation of BSS in Figure 8 (right). Here, most of the registered BSS are distributed along with these two options. For the following rounds, larger EPR of at least two hours were required. However, there are still projects following an EPR of one hour, which will probably change their EPR design to meet the requirements.

After high award volumes in April 2022, there was a sharp drop in the December round in which only one operator submitted a bid. This was again due to a change in regulation. Whereas a fixed market premium was tendered until April 2022, a flexible market premium was introduced in December 2022, which reduces the rate of return and led to the drop in bid submissions. The market premium is paid for a period of 20 years, which must be guaranteed as the lifetime of the BSS and PV. As this is quite long for a BSS, this requirement led to some BSS being operated only little to minimize aging.

The areas of application of the BSS are not prescribed in the innovation auction. However, the stipulation that the BSS may only be charged by the directly connected renewable energy power plants results in many restrictions. Any bidirectional use case (charge and discharge to the grid) such as the participation in the FCR market is thus excluded, as is the provision of negative aFRR and the purchase of energy on the electricity market. Under the current regulations, the only conceivable areas of application are the deferred sale of energy from the renewable energy power plant on the electricity market, the provision of positive aFRR, and the reduction of forecast deviations in renewable energy feed-in in the form of peak shaving and steady feed-in. As the BSS is only allowed to charge from the RE system, the installed resources are not used over large parts of the year, when the sun is not shining during nights and with low production in winter months. If the regulation changed and BSS were allowed to charge from the grid and to participate in the ancillary service and spot markets, this would be a good flexibility addition for the energy transition.

### III.D. Electric vehicle market in Germany

Battery electric vehicles (BEV) and plug-in hybrid electric vehicles (PHEV) are being adopted in the mass market. As a consequence, their battery energy nowadays greatly exceeds the stationary market. In 2022, the EV stock grew by around 693,000 EV (see Figure 9) with a battery energy of 27 GWh (see Figure 10), corresponding to a growth of 34% in comparison to the battery energy additions in 2021. While BEV and PHEV were on par over the last years in terms of numbers, BEV are taking over the market as the transition technology PHEV begins to fade out.

In terms of numbers, BEV registrations grew by 28% and PHEV by only 4% (see Figure 9).

By the end of 2022, an estimated 1,878,000 EV with a battery energy of 64.64 GWh were operated in Germany. The majority of the battery energy (53.26 GWh) was installed in slightly over one million BEV and 11.38 GWh in 0.86 million PHEV (see Figure 10). This puts vehicle batteries officially at the top of energy storage in Germany as they exceed the approximately 39 GWh of national pumped hydro storage power plants currently in operation [64] by a factor of 1.7.

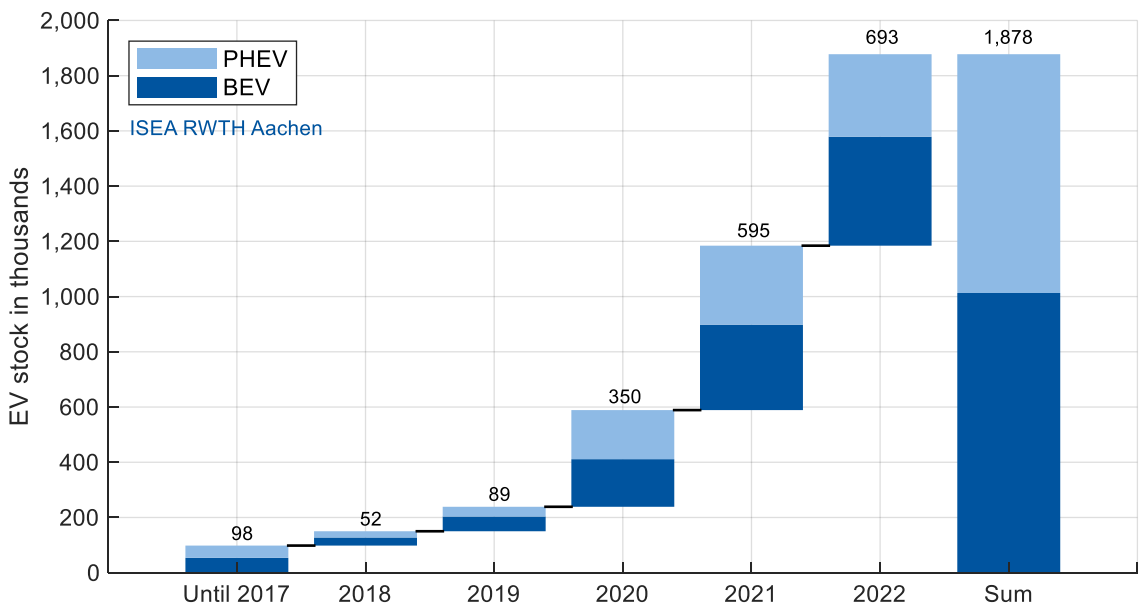

*Figure 9. Estimated number of EV stock in Germany based on own analyses of FMTA DB [12].*

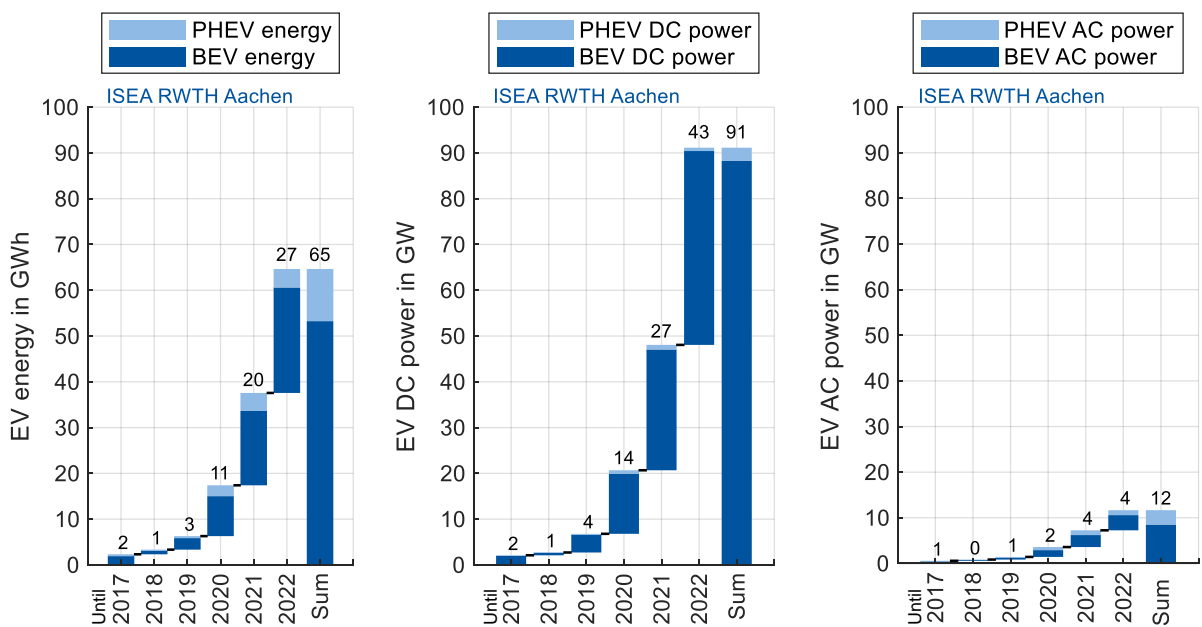

*Figure 10. Estimated EV battery energy (left), DC charging power (middle), and AC battery charging power (right) based on combination of EV registrations in FMTA DB [12] and vehicle data sheets from ADAC DB [37]. Note that 2022 values might change slightly as the comprehensive EV stock dataset "FZ6" has not been published at time of writing.*

New peaks were also reached in total charging power. All DC connections on vehicles combined had 91.14 GW and AC connections had about 11.67 GW. The DC power already exceeded the maximum load of Germany in 2022 (around 80 GW [65]). However, both DC and AC charging values are to be understood as theoretical, since only a fraction of the vehicles are charging at the same time, and often with much lower power. For comparison: The sum of all national public charging stations was 2.47 GW by the end of 2022 (see section III.F).

Breaking the total battery energy and power down to the individual vehicles, clear trends can be seen in new cars over the last few years. While we estimate that an average of 47.1 kWh of battery energy per BEV was installed in 2019, this figure has risen by 24% to 58.4 kWh in 2022 (see appendix, Figure 25). The typical consumption varies between 15 kWh to 25 kWh per 100 km (see appendix, Figure 28) and average ranges are around 300 km to 450 km (see appendix, Figure 28). For a typical daily commute of around 40 km in Germany [66], this means charging will only be necessary once a week on average. The average battery energy for plug-in hybrids is significantly lower at an average of 10.2 kWh in 2022. The energy density for the whole battery pack is around 150 Wh/kg (see appendix, Figure 29).

When it comes to charging power, the differences between AC and DC charging are very clear. More and more vehicles have a DC interface, and these interfaces allow high power charging. This results in an average charging power of 107 kW per new BEV in 2022 (see appendix, Figure 25 and Figure 27), with vehicles without a DC connection being counted as 0 kW, thus lowering the average. If only vehicles equipped with a DC connector are considered, the average DC power rises to 116 kW in 2022 (102 kW in 2021). Charging times of the BEV models average from around 30 min to 45 min (see appendix, Figure 29) to recharge from 0% to 80% SOC. However, high powers cannot be applied for the whole charging process and typically decrease continuously with increasing SOC, until they drop sharply from 80% SOC upwards due to the constant voltage phase.

For AC charging connectors, on the other hand, there has been little movement and the value has hovered between 8 kW and 9 kW since 2019. This value is primarily due to the fact that although approximately 97% of the vehicles have a Type 2 charging connector, they cannot always also be charged via all three installed phases. Also, often only 16 A per phase are transmitted and not 32 A or even the theoretically possible 64 A. This results in a charging power of 11 kW for most vehicles and 3.7 kW or 7.4 kW for some vehicles. This design decision is understandable, considering that residential charging stations typically can only provide 11 kW. If the power electronics in the vehicle were sized for 22 kW, then the charger would only operate at partial load producing greater losses during charging at low power. About 41% of the vehicle stock consisted of models that are only sold with a single-phase connection. Another 14% are sold solely with a three-phase connection. Note that the existing data is insufficient to differentiate between a one-, two-, or three-phase connection as models are sold with varying specifications. DC fast charging connectors are available in approximately 45% of vehicles with CCS being the most prominent option representing 79% of fast charging connectors. Additional 13% of DC fast charging connectors are either CCS or Tesla Superchargers and cannot be split up precisely between those options. The highest share is likely to be CCS as Tesla announced to use CCS by the end of 2018 [67].

Discharging the vehicle into the public grid, called vehicle-to-grid [68], is currently not possible from a regulatory point of view in Germany. There are individual vehicle manufacturers and CS operators that offer or plan to enable bidirectional charging equipment and vehicles [69, 70] with the recently published standard ISO 15118-20 [71]. Due to the regulatory challenges for vehicle-to-grid, a scenario in which the vehicle acts similar to a HSS (without grid feed-in from the battery) in form of vehicle-to-home [72] is most likely to start this year.

### III.E. Prices

For many years, lithium-ion prices have been falling worldwide. However, in 2022, the world market battery prices increased by 7% to 151 $/kWh (144 €/kWh) according to a survey by Bloomberg New Energy Finance (BNEF) [73].

#### III.E.1. Stationary storage

In general, prices are not transparently available for a number of reasons. Among these are (1) prices vary largely depending on the region [73]. (2) It is not always clear what is included and values can reach from pure prices for the hardware up to turnkey prices inclusive ground and value-added taxes (VAT) covering large price ranges [2]. In this regard, some sources speak of prices while others of costs, both meaning probably the same [74]. (3) Information provided by manufacturers have to be seen as indicative as the projects might have completely different circumstances [2]. The incentive for manufacturers to state low prices is questionable as this would set expectations for customers. (4) Further, demand and supply are quite dynamic, which leads to fluctuating prices to balance this situation.

For HSS, we have gathered end-customer prices (see Figure 11) from around 30,000 HSS until 2021 in a German-wide subsidy program and a subsidy program in Baden-Württemberg (BW) (see Table 3). The prices in BW for average-sized HSS from 5 kWh to 15 kWh (see Figure 19) were slightly higher than in our German-wide monitoring probably due to a general higher price level in southern parts of Germany or higher subsidies [36]. From 2013 to 2021, HSS prices fell by 63% from 2,500 €/kWh to 920 €/kWh. However, in 2022, we estimate a price increase of around 30% after discussions with partners from industry, leading to an estimated mean price of around 1,200 €/kWh. The reason for the price increase is the extremely high demand, which has continued to grow due to high energy prices and the war in Ukraine. Installers are at the limits of their installation capacities and many people have to wait more than half a year for an installation date of PV and HSS. However, the prices vary largely and we have seen prices ranging from 700 €/kWh up to 2,300 €/kWh in single offers analyzed in 2022.

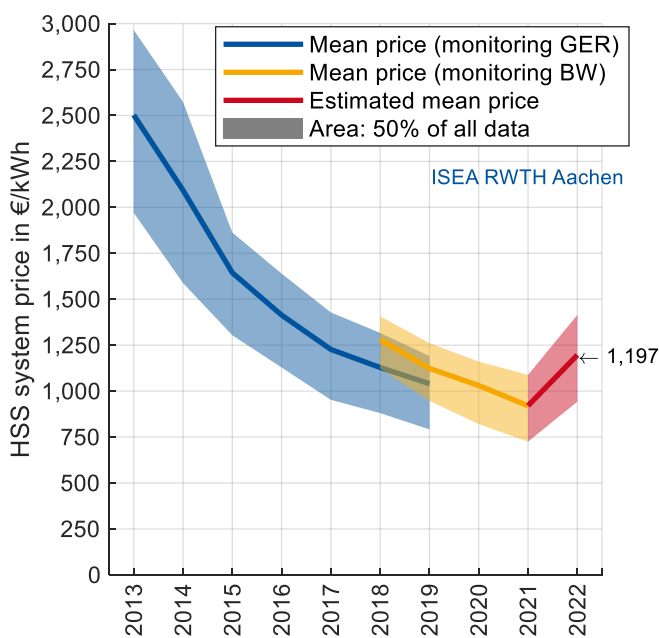

*Figure 11. Price development of 30,000 HSS per usable energy according to data in MONITORING DB and an estimate of 30% increase from 2021 to 2022 based on bilateral exchange with industry. Values adjusted for inflation (base year: 2022) and without VAT.*

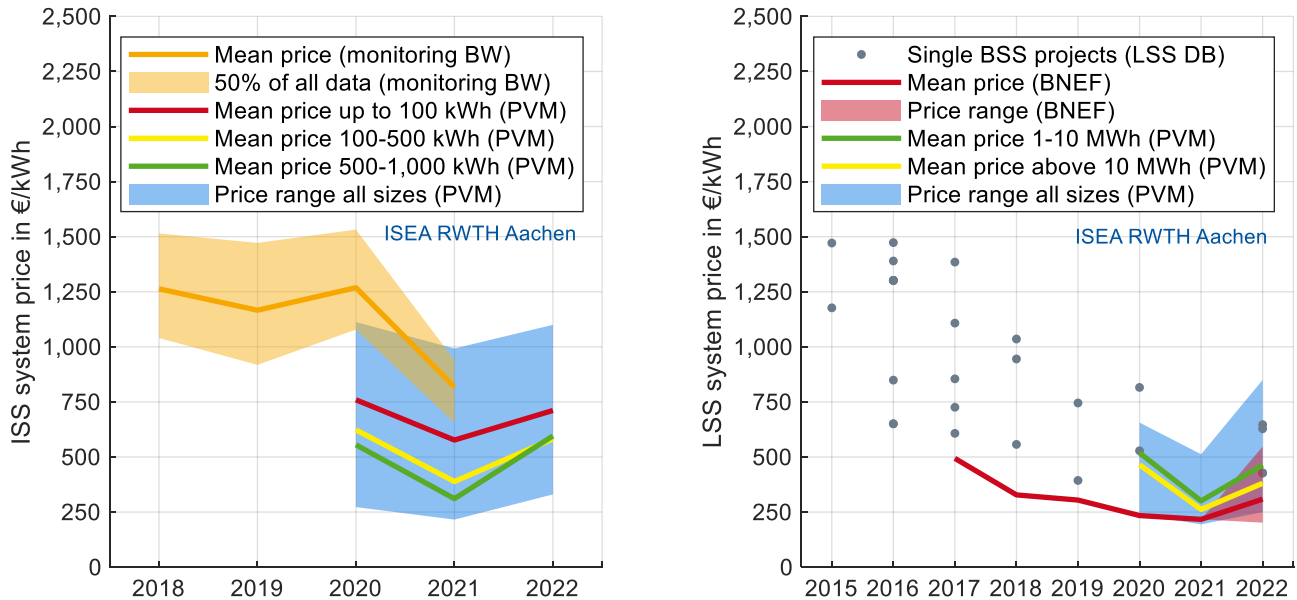

*Figure 12. Price development of ISS (left) and LSS (right) according to data in MONITORING DB, extended LSS DB [35], own analyses of pv magazine (PVM) database [75], and Bloomberg New Energy Finance (BNEF) [79]. Monitoring DB and PVM adjusted for inflation (base year: 2022), BENF values (already adjusted for inflation by BNEF) converted from USD to EUR. Prices per usable energy and exclusive VAT except for unknown price components from LSS DB [35].*

For ISS, we have data from the subsidy program in BW until 2021 (mostly below 100 kWh), where average prices were 820 €/kWh (see Figure 12). The pv magazine (PVM) collects indicative prices from manufacturers for ISS and LSS in a database [75] at the very beginning of the year which is why we assign the prices rather to the previous year. Our analysis changes slightly from the reported values by PVM in [76–78] as we show the mean prices for 2022 and not the median prices. In 2021, the mean prices range from 215 €/kWh to 990 €/kWh depending on the ISS energy and manufacturer. From 2021 to 2022, mean values for ISS up to 100 kWh increased by 23% from 575 €/kWh to 710 €/kWh, for ISS from 100 kWh to 500 kWh by 50% from 390 €/kWh to 580 €/kWh, and for ISS from 500 kWh to 1,000 kWh by 92% from 310 €/kWh to 595 €/kWh. However, it should be noted that not exactly the same manufactures provided prices for 2021 and 2022.

Single LSS project prices are included in the LSS DB for the years 2015 until 2020 and we also scanned further press releases for 2021 and 2022, but it is not always clear what exactly the prices take into account. They depict a price decrease from the average of 2017 (1,080 €/kWh) of around 50% until 2022 (532 €/kWh). From 2021 to 2022, the prices from pv magazine show a price increase of 54% for LSS between 1 MWh and 10 MWh from 300 €/kWh to 464 €/MWh. LSS above 10 MWh showed a price increase of 45% from around 250 €/kWh to 380 €/kWh for the same time period. For 2022, all values range from 250 €/kWh to 850 €/kWh depending on the energy and manufacturer. The 2022 mean prices from pv magazine are supported by the mean price of 310 €/kWh reported by BNEF for a four-hour turnkey LSS [79]. The authors point out, that prices depend largely on region, energy, and system design and cover a range from 292 €/kWh to 548 €/kWh [79]. The identified price increase of 27% from 2021 to 2022 in USD [79] corresponds to an increase of 42% when converted to EUR.

### III.E.2. Battery electric vehicles

Figure 13 shows the BEV price over the range according to our analysis of ADAC DB [37], first presented by us in [80]. PHEV are neglected for the following analyses.

Over the past few years, manufacturers have greatly expanded the product range of available BEV models while closing it from two sides. Initially (2013-2017), there were only small, lower medium and luxury BEV available while medium and upper medium BEV started from 2018 and 2020, respectively (see Figure 14 and appendix, Table 4 for vehicle examples). Most models offer ranges between 300 km and 550 km and are priced between 25,000 € and 75,000 € (see Figure 13). It can be observed that in general there is a positive correlation between range and price for the different segments, but prices in the luxury segment are clearly above all other segments.

Absolute vehicle prices have increased faster than inflation as an analysis of the ADAC [81] and the data in ADAC DB shows. However, BEV manufacturers can purchase batteries cheaper than stationary BSS manufacturers as they order higher volumes [73]. As EV can also be operated in bidirectional use cases, we introduce the "specific BEV system price" (BEV price divided by battery energy) to compare the mobile storage system "BEV" with stationary storage systems. Figure 14 reveals that specific BEV system prices have decreased for some vehicle classes while battery energy has increased (see appendix, Figure 26).

From year 2014 to 2022, specific BEV prices have fallen by 31% for small BEV and by 52% for lower medium BEV. Luxury car prices largely depend on the manufacturer and model and no clear trend can be observed.

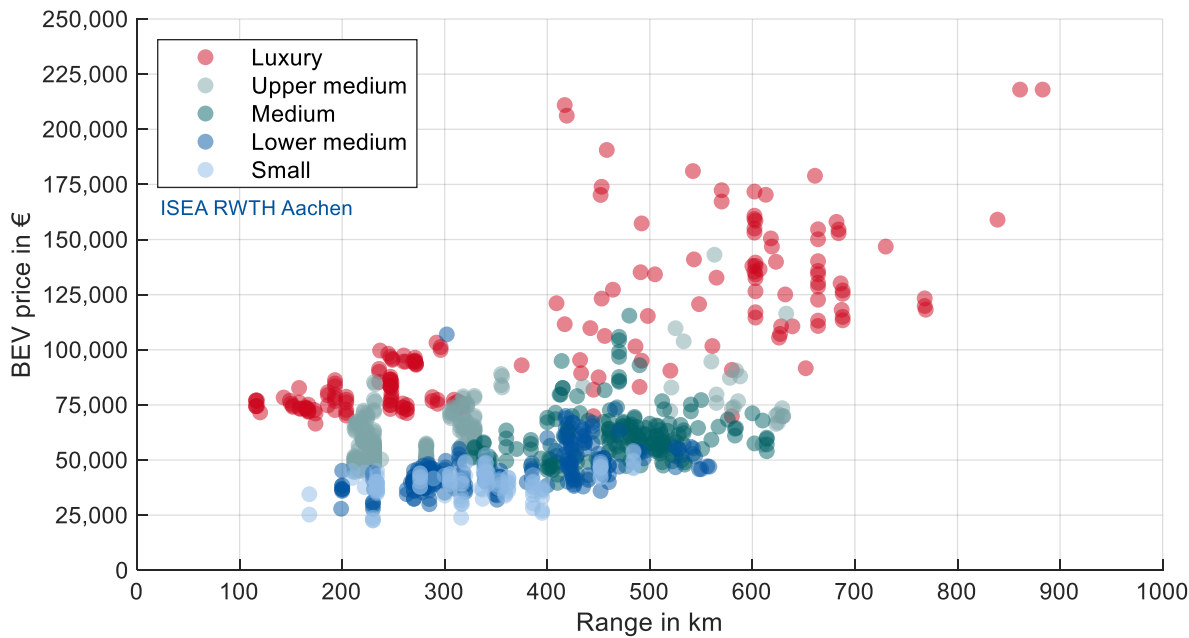

Figure 13. Development of BEV prices over the driving range. Data from ADAC DB [37] and prices adjusted for inflation. The shown values are base prices and represent the lower price limit as extra equipment is not included.

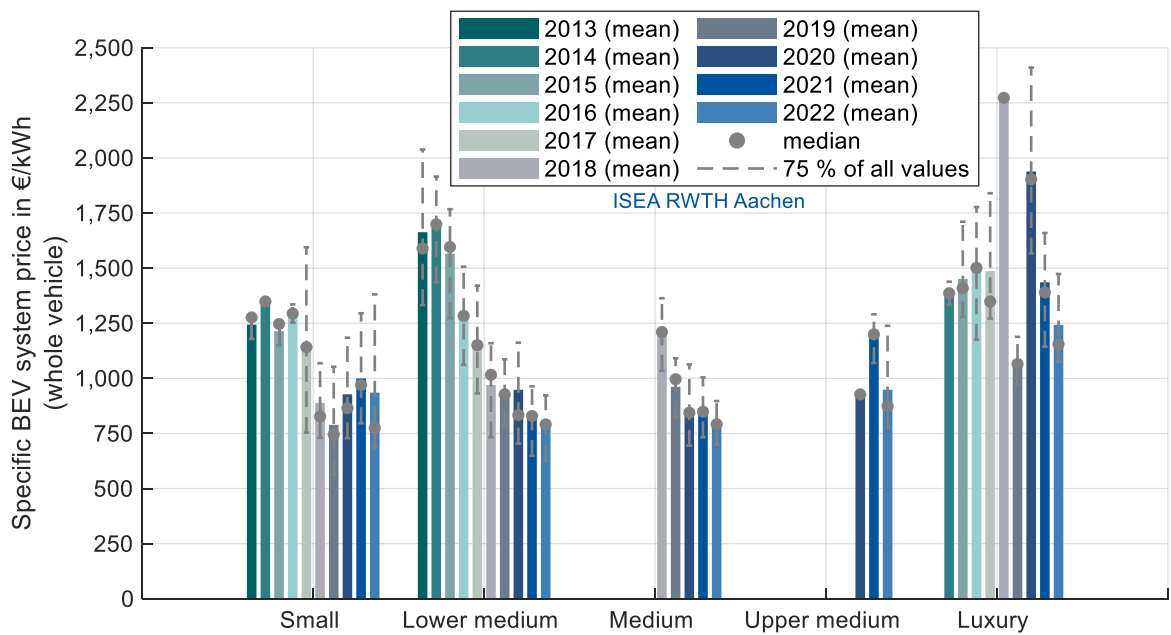

Figure 14. Development of specific BEV prices per usable battery energy for different vehicle classes over time based on own analysis of ADAC DB [37]. Prices adjusted for inflation. The shown values are base prices and represent the lower price limit as extra equipment is not included. Prices are not weighted by sales volume and show the BEV models that started in the specified year.

For the same period, world market battery pack prices decreased by around 77% [73]. Further, the battery makes up about 30% to 40% of the EV price [82].

Therefore, it is clear that falling battery prices were one of the major drivers for falling specific BEV system prices. However, although world market battery prices have risen from 2021 to 2022, specific BEV system prices decreased by 6% for small, by 2% for lower medium, by 8% for medium, by 20% for upper medium and by 13% for luxury cars. Small cars had an average specific BEV system price of 930 €/kWh in 2022. While the specific BEV system prices for the lower medium and medium cars were around 800 €/kWh, they were around 1,240 €/kWh for luxury cars in 2022. These contrary developments can possibly be attributed to economy of scale as production capacities increase. However, EV pricing strategies also depend on many other factors. Since electromobility is still in its infancy, further cost reductions can be expected.

Comparing the specific BEV system prices it becomes clear how much economic potential bidirectional EV use cases have. Even though the prices are for the whole BEV, which is technically more complex than stationary storage, most prices are lower than HSS prices which makes vehicle-to-home quite attractive as people can also "drive" their storage. Depending on the size, ISS have comparable or lower prices and LSS show mostly lower mean prices (see section III.E.1).

### III.F. Charging infrastructure in Germany

In 2022, 21,000 new public EV charging points (CP) were installed (see Figure 15). Thus, EV could charge at around 80,500 CP at roughly half the number of charging stations (CS) at the end of 2022. This corresponds to about two CP per CS. Most CP are in the range of 15 kW to 22 kW, with about 66% of the stock. CP with medium powers (22 kW to 149 kW) take up only 8% of the installations and fast charging (149 kW to 299 kW) and ultra-fast charging (above 299 kW) 5% and 4%, respectively. The cumulative charging power is therefore only 2.47 GW and represents a small fraction of the theoretical EV power shown in section III.D.

The annual additions of CP cannot keep up with the accelerating growth in EV sales. In 2022, the growth with respect to the number of CP and the previous year was 24%. However, the segments contribute differently. With regard to 2021, new CP installations up to 3.7 kW decreased by 51% and CP from 49 kW to 59 kW by 16%. For CP from 22 kW to 49 kW, the number of decommissioned CP was even higher than the number of new installations. In contrast, CP installations from 3.7 kW to 15 kW increased by 10%, CP from 59 kW to 149 kW by 35%, CP from 149 kW to 299 kW by 78%, and CP above 299 kW by 30%. The growth in CP above 150 kW can be explained by the better economics for fast charging CP due to a higher energy throughput as shown in [83, 84].

The relatively small growth of CP over the last years led to an increasing number of EV per CP. While in 2017 there were only 9.1 EV/CP, this number has risen to 23.3 EV/CP by 2022 (see Figure 30). When only BEV are considered, the number has grown from 5.0 BEV/CP (2017) to 12.6 BEV/CP. Nevertheless, fast charging installations make progress: for a power above 150 kW the value of BEV per fast charging point (FCP) has been decreasing since 2017 from 1,456 BEV/FCP to 144 BEV/FEC in 2022. To push public charging infrastructure, the Federal Ministry for Digital and Transport runs several subsidy programs. The most prominent one is the "Deutschlandnetz" [85], a German-wide grid of 1,100 fast charging stations with planned public spending of 1.9 billion Euros. This grid should allow fast chargers to be reached all over Germany in less than 10 min. To achieve this, the government defined areas at central traffic intersections. Out of these, 200 are built at parking locations along highways owned by the Federal Government and 900 are located in so-called "search areas" with a radius of approximately 2 km [85]. Each charging station must have 4 to 16 CP depending on the traffic volume. The locations are tendered Europe-wide [86] with the winner being awarded based on an economic basis with limits on the maximum customer price in discussion. Among several other requirements, each bidder must ensure at least the following for new charging stations [86]:

- User comfort ensured through roof, gastronomic outlet, sanitary facilities, and wheelchair-friendliness
- Minimum power of 300 kW DC per CP offered at least via CCS. If the station is fully occupied, at least 200 kW must be available for each CP
- Support for 400 V and 800 V systems
- Access through RFID-card, App, Plug-&-Charge (once available), and ad-hoc charging without registration through an NFC card pad

### III.G. Production capacities in Germany

Recently, around 2 TWh yearly production capacity of so-called "gigafactories" for batteries have been announced by various manufacturers in Europe with Germany leading the number of planned factories [87, 88]. To keep track of these developments, some institutions, including us, have been gathering the announcements mostly from news websites and press releases.

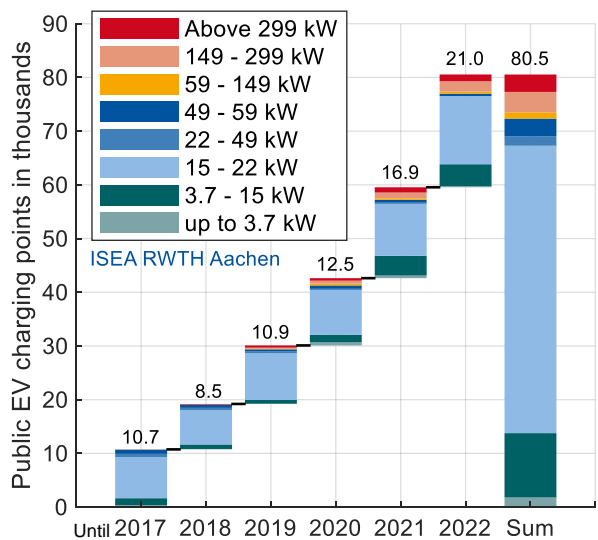

*Figure 15. Estimated number of public EV charging points in Germany based on registrations at FNA in CHARGE DB [12].*

We estimate that the current announcements from the various companies add up to around 364 GWh/a for Germany as of March 2023 (see Figure 16). This makes clear how large the developments of the next years have to be. Depending on the source, the announcements range from 227 GWh/a to over 546 GWh/a.

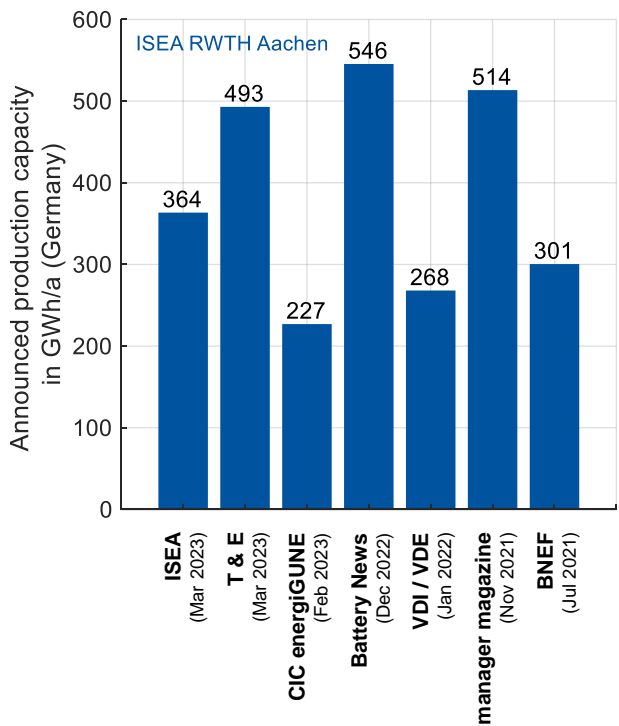

*Figure 16: Announced production capacities in Germany based on own research and selected other public sources. These information change frequently and updates should be checked. Sources: ISEA [89–105], Transport & Environment (T&E) [87], CIC energiGUNE [106], BATTERY-NEWS.DE [88], manager magazin [107], BNEF [108], VDI / VDE [109].*

If these plans are realized, Germany would probably export batteries mostly in form of EV. The information change quickly due to new press releases. We expect further dynamic changes, as companies might shift the entire or parts of production to the US due to the US Inflation Reduction Act. A recent report by Transport & Environment [87] outlines the current risk of the European battery production in regard to delay, scale down, or cancellation very clearly. The authors estimate 16% of European gigafactories to have a high risk, 52% medium risk and 32% low risk. For Germany, these values are 18% (high), 62% (medium), and 20% (low), respectively. The proposed measures to handle the risk are a European response to US subsidies and faster regulatory approvals [87].

## IV. CONCLUSION AND OUTLOOK

Figure 17 summarizes the key information of the paper while focusing on the installed battery energy. All in all, the stationary and mobile BSS markets are growing strongly. In 2022, an additional battery energy of 30 GWh was installed, of which only 9% accounted to stationary BSS and 91% to EV. By the end of 2022, all markets had a cumulative battery energy of about 72 GWh. Within the stationary market (7 GWh), HSS are leading with 78% of the newly installed battery energy in 2022 and 79% (5.49 GWh) of the stock, while ISS account for 4% (0.27 GWh) and LSS for 17% (1.2 GWh) of the stock (see appendix, Figure 24). In the EV market, BEV and PHEV had a balanced share in terms of numbers until 2021, but 2022 brought more BEV. In terms of battery energy, BEV account for 85% of the new registrations and 82% of the EV stock. By the end of 2022, battery storage had nearly twice the energy of the national pumped hydro storage power plants of 39 GWh [64]. This shows the enormous flexibility potential of battery storage in our energy system. However, this potential has to be unlocked through vehicle-to-grid technologies with bidirectional EV and charging stations, which is only at the beginning of market realization [110].

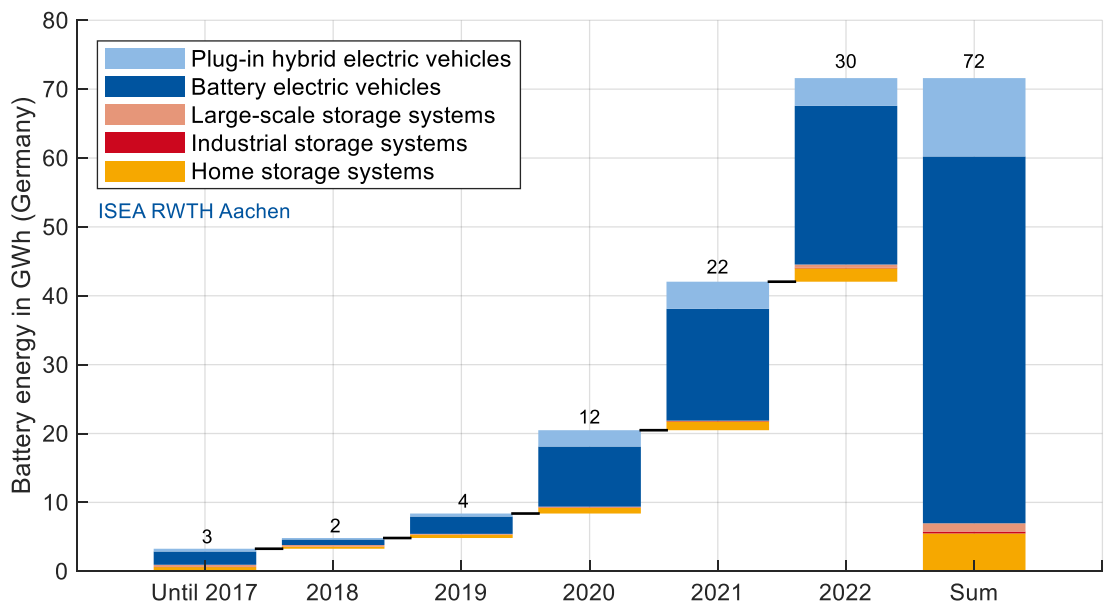

*Figure 17. Estimated stationary and mobile battery storage market in Germany.*

At the moment, EV flexibility could be provided through smart charging strategies. In the following, the several markets are presented.

### IV.A.  Home storage systems

The HSS market continues to grow hand in hand with the PV market. In 2022, 220,000 HSS with a battery energy of 1.94 GWh and a power of 1.16 GW were installed in Germany, showing a market growth of 52% in terms of installed energy. By the end of 2022, 650,000 HSS with an energy of 5.49 GWh and a power of 3.1 GW were installed.

The Federal Network Agency approved the grid development plan of the transmissions system operators that currently plan with 97.7 GW to 113.4 GW for 2045, probably also including ISS [111].

The PV market will most likely continue to grow due to the ambitious plans of the new government to reach yearly PV installations of 20 GW/a in comparison to the 7.5 GW/a in 2022 [112]. In addition to the joint installations with new PV systems, more and more old PV systems that run out of the feed-in tariff will become available in the coming years. Further, higher energy prices and residential sector coupling are driving the installations, as more and more HSS are operated in combination with an EV or a heat pump [36]. While EV and HSS push each other right know, it stays unclear what effect the vehicle-to-home application of EV will have on the market. For some people EV will not be at home when the sun shines and HSS will be needed, for others (home office, second car) EV could serve similar to HSS reducing their sales.

### IV.B.  Industrial storage systems

The ISS market grew in 2022 with 1,200 new ISS with a battery energy of 0.08 GWh and a power of 0.04 GW showing a growth of 24% in terms of battery energy. By the end of 2022, about 3,900 ISS with a battery energy of 0.27 GWh and a power of 0.14 GW were installed in Germany.

We expect much stronger growth in the coming years. Everything points in this direction: the German Climate Protection Act stipulates a reduction in annual emissions volumes in industry of around 35% from 2020 to 2030 [113], and electricity prices for industry are higher than ever [114]. However, this is also accompanied by a much higher fluctuation of the exchange electricity price during a day. If storage facilities can be managed wisely, this currently results in much higher margins, which contribute significantly to the profitability of storage operations. ISS play a critical role for $CO_2$-free industrial sites by decoupling generation and consumption over time. The versatile applications include, for example, increasing solar self-consumption, peak shaving for grid fee reduction, uninterruptible power supply or the integration of renewable energies and electric vehicles. In particular, some of the planned fast charging stations will also have to be operated with buffer storage, which will give the market a boost. Here, the ISS enable charging stations to provide higher power at short notice than the grid connection would allow. The ISS are then recharged at times when the charging stations are not very busy.

### IV.C.  Large-scale storage systems

In 2022, a record of 47 LSS with a battery energy of 0.47 GWh and a power of 0.43 GW were installed in Germany, showing an increase of 910% in terms of battery energy. By the end of 2022, 149 LSS with a cumulative battery energy of 1.2 GWh and a power of 1.07 GW were installed. For the next two to three years, there is with 1.41 GWh already more LSS energy announced than the current stock by the end of 2022.

LSS power with a focus on FCR already exceeds the national FCR market, but new application areas in form of the renewable energy (RE) integration, industrial energy supply, arbitrage trading, automatic frequency restoration reserve, grid boosters, as well as multi-use operation emerge.

As more and more conventional power plants with their rotating masses disappear from the market, there will be a high demand for synthetic inertia in the future. Batteries are capable of this in principle, but the extremely fast response must be ensured by the power electronics [115]. The inverters used must operate in a grid-forming and voltage-imprinting manner in the interconnected grid, emulating the characteristics of synchronous generators in the event of faults.

The Federal Network Agency approved the grid development plan of the transmissions system operators that currently plan with an LSS power of 43.3 GW to 54.5 GW for 2045 [111].

### IV.D.  Battery electric vehicles

The market for electric vehicles (EV) grew with 693,000 EV new registrations in 2022 and a battery energy of 27 GWh, a DC charging power of 43 GW, and an AC charging power of 4.5 GW by 34% in terms of battery energy. By the end of 2022, 1,878,000 EV with a battery energy of 65 GWh, a DC charging power of 91 GW, and an AC charging power of 12 GW were operated in Germany.

The EV market is and will likely remain highly dynamic. Virtually all large vehicle manufacturers have an EV strategy and a growing number of companies have announced an end of sale for conventional internal combustion engine vehicles. Examples are Volvo (2030), Ford (2030, in Europe), Jaguar Land Rover (2025), General Motors (2035), and Audi (2035) [116].

The new German government has announced the goal of reaching 15 million EV on the road by 2030 as well as 1 million available charging points [117]. It is further likely that the trend of larger battery energies and higher charging power per vehicle will continue in an attempt to make the driving experience of an electric vehicle comparable to an internal combustion engine vehicle. Given these two trends, the battery energy of the entire vehicle fleet in the range of hundreds or thousands of GWh seem realistic. If an average energy of 70 kWh per EV is assumed for the 2030 fleet, there will be 1.05 TWh on the road. DC charging power with an assumed average of 150 kW per EV would add up to 2.25 TW on the EV side although the sum of charging station power will probably be lower and only a fraction of EV will charge simultaneously. Given intelligent charging algorithms, this flexibility is likely to severely disrupt the current storage market. If integrated properly into the grid, these vehicles could for instance largely complement the stationary BSS in ancillary services and participate in arbitrage trading [118]. The association of European TSOs (ENTSO-E) has already published a position paper about the role of EV in the energy system [119]. To achieve this integration, the industry is working on innovative charging solutions such as bidirectional charging, simple-to-use authentication and payment at public charging stations, and intelligent load management.

*IV.E. Prices*

While inflation-adjusted world battery prices increased for the first time by 7% from 2021 to 2022 (144 €/kWh) [73], stationary battery system prices increased way more. With price increases estimated at 30%, HSS averaged 1,200 €/kWh, ISS with price increases between 23% and 92% depending on size averaged between 580 €/kWh and 710 €/kWh, and LSS with increases from 42% to 54% averaged between 310 €/kWh and 465 €/kWh. In contrast, specific BEV system prices (whole BEV price divided by battery energy) decreased, probably due to economy of scale, depending on vehicle class. Specific BEV system prices for small cars fell by 6% to 930 €/kWh, for lower medium cars by 2% to 800 €/kWh, and medium cars by 8% to 800 €/kWh. The fact that BEV system prices are lower than HSS prices and only slightly higher than ISS prices shows the low purchase prices for automotive manufacturers and makes vehicle-to-home, vehicle-to-building, or vehicle-to-grid use cases quite attractive.

*IV.F. Charging infrastructure*

Charging point (CP) installations need to accelerate to keep up with the boom in EV. The factor of EV per CP has risen from 2017 to 2022 from 9 to 23; if only BEV are considered, from 5 to 13 within the same period. Nevertheless, fast charging above 150 kW makes progress as the number of BEV per fast CP decreased from 1,456 (2017) to 144 (2022). In 2022, 21,000 new CP were installed in Germany, showing a growth of 24% in terms of numbers and the previous year. Especially the fast CP above 150 kW grew significantly (55%) as they show higher occupancies and better economics. By the end of 2022, about 80,500 public CP were installed, of which 66% have a power between 15 kW and 22 kW. To increase installations, the federal government is pushing fast charging stations in particular with the fast charging law and the "Deutschlandnetz".

*IV.G. Production capacities*

Many companies have announced so-called "gigafactories" to be built in Europe with a focus on Germany. The collected announcements range from 227 GWh/a to 546 GWh/a. However, many projects bear the risk of a delay, slow down, or cancellation as the US Reduction Inflation Act could pull production overseas and Europe has no answer until now. These announcements change frequently and need to be checked regularly.

*IV.H. Future market tracking*

The market is very dynamic and numbers are therefore soon not up to date any more. We could show that, for new installations, non-reported BSS to the MASTR DB of the Federal Network Agency are only a few percent and most of them will be registered belated in the coming months. To provide the readers from now on with the latest registration figures, we present the stationary battery storage registrations in MASTR DB on www.battery-charts.de [32] as supplementary material to the paper. Further, the reliable EV registrations at the German Federal Motor Transport Authority are depicted on www.mobility-charts.de [33], as presented in [3]. Both websites offer interactive data exploration.

## V. Acknowledgement

Parts of the results were obtained within the research project "Betterbat" (funding number 03XP0362B), funded by the Federal Ministry of Education and Research (BMBF). Other parts were obtained in context of the research projects "Speichermonitoring BW 2.0" (funding number L75 21120) and "Speichermonitoring BW" (funding number L75 18006), both funded by the Ministry of Environment, Climate Protection and the Energy Sector, Baden Württemberg (UMBW), as well as in the research projects "WMEP PV-Speicher" (funding number 0325666), "WMEP PV-Speicher 2.0 (KfW 275)" (funding number 03ET6117)", funded by the German Federal Ministry for Economic Affairs and Climate Action (BMWK). The authors take full and sole responsibility for the content of this paper.

## VI. Author Contributions

**Jan Figgener**: Conceptualization, Methodology, Software, Formal analysis, Investigation, Data Curation, Writing – Original Draft, Visualization, Funding Acquisition. **Christopher Hecht**: Conceptualization, Methodology, Software, Formal analysis, Investigation, Data Curation, Writing – Original Draft. **David Haberschusz**: Conceptualization, Methodology, Data Curation, Writing – Review & Editing. **Jakob Bors**: Data Curation, Software, Writing – Review & Editing. **Kai Gerd Spreuer**: Data Curation, Software. **Kai-Philipp Kairies**: Writing – Review & Editing. **Peter Stenzel**: Writing – Review & Editing. **Dirk Uwe Sauer**: Conceptualization, Resources, Writing – Review & Editing, Supervision, Funding Acquisition.

VII. APPENDIX

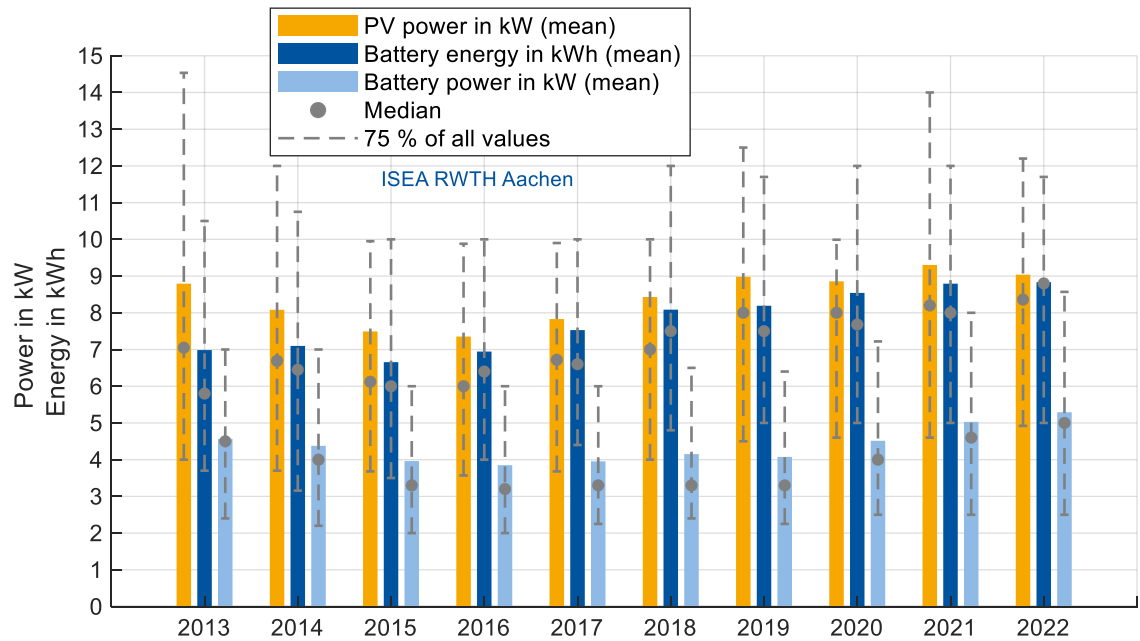

*Figure 18. Mean PV power (2 kW to 30 kW), and HSS battery energy and inverter power in Germany based on own analyses of MASTR DB [34].*

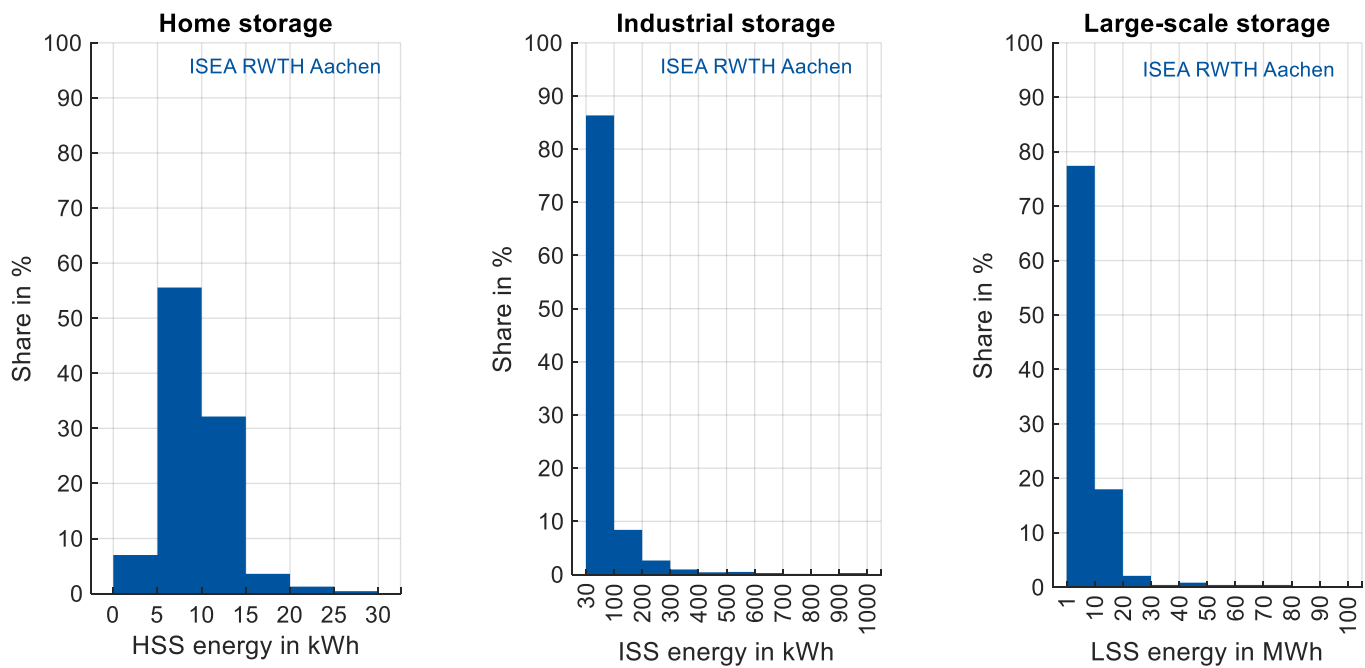

*Figure 19. Share of BSS number in energy classes of HSS (left), ISS (middle), and LSS (right) in Germany based on own analyses of BSS in operation and registered in MASTR DB [34].*

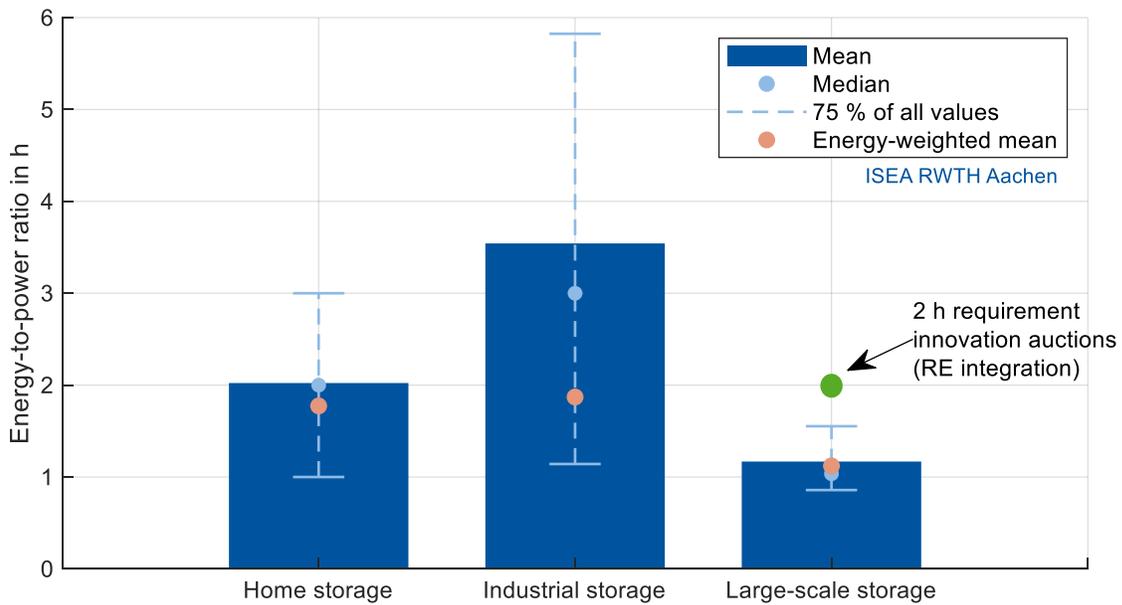

*Figure 20. Energy-to-power ratio distribution of stationary battery storage systems in Germany based on own analyses of BSS in operation and registered in MASTR DB [34].*

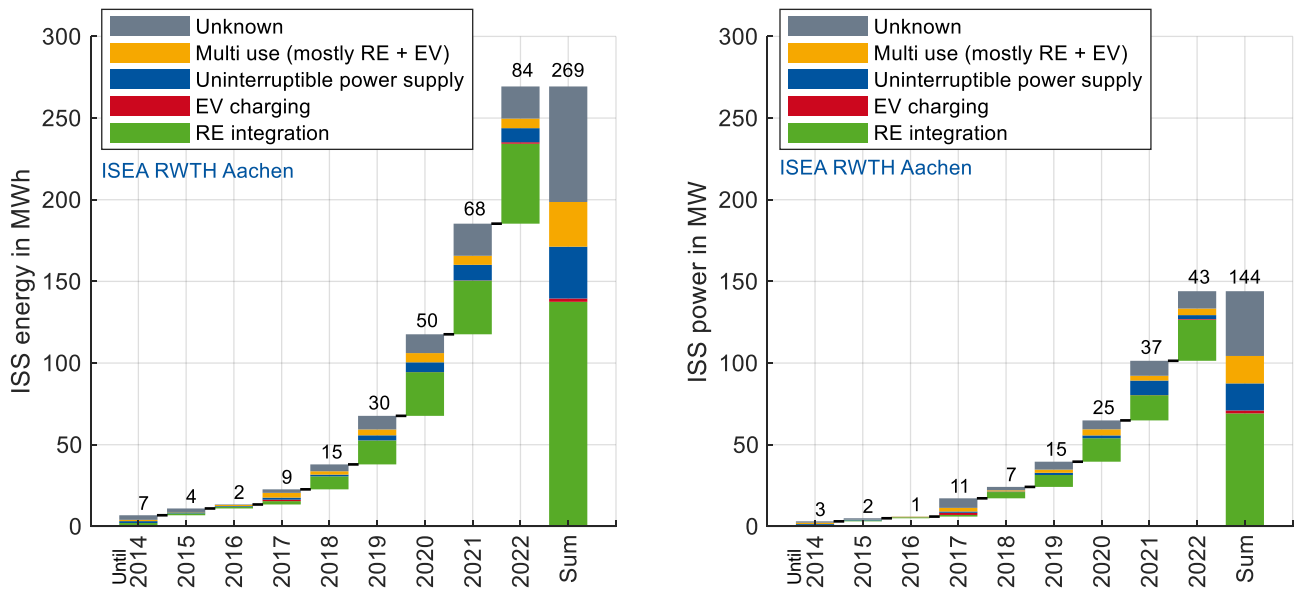

*Figure 21. Development of ISS battery energy (left) and inverter power (right) by use case in Germany based on own analyses of MASTR DB [34] and further research.*

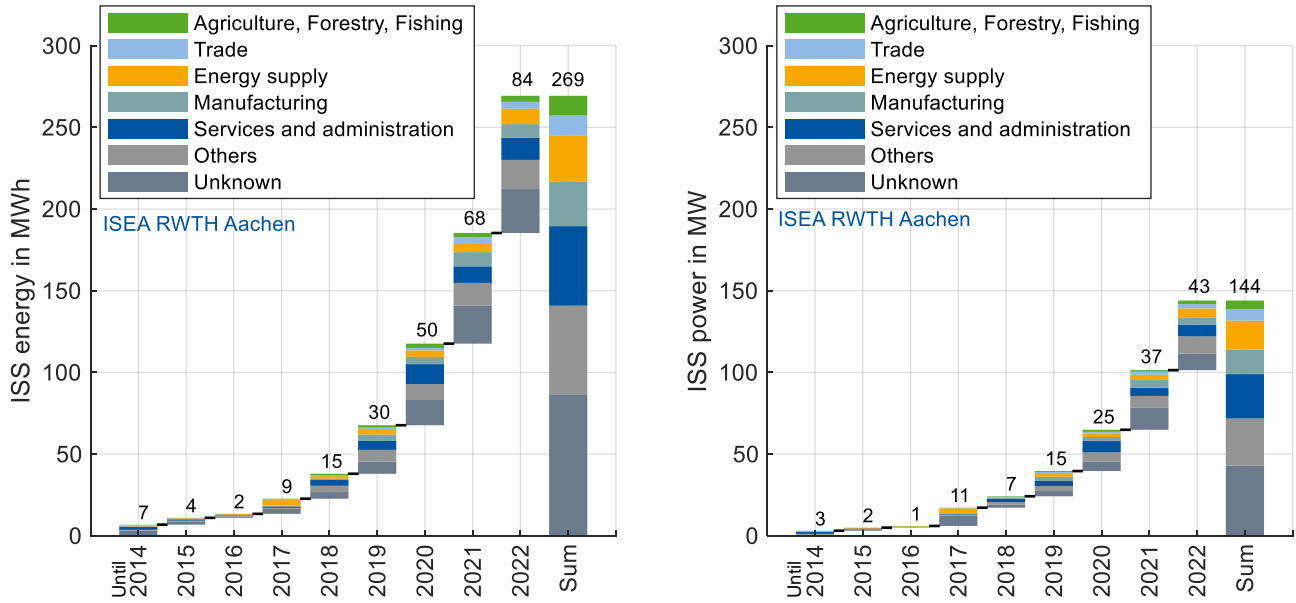

*Figure 22. Development of ISS battery energy (left) and inverter power (right) by economic sector in Germany based on own analyses of MASTR DB [34] and further research.*

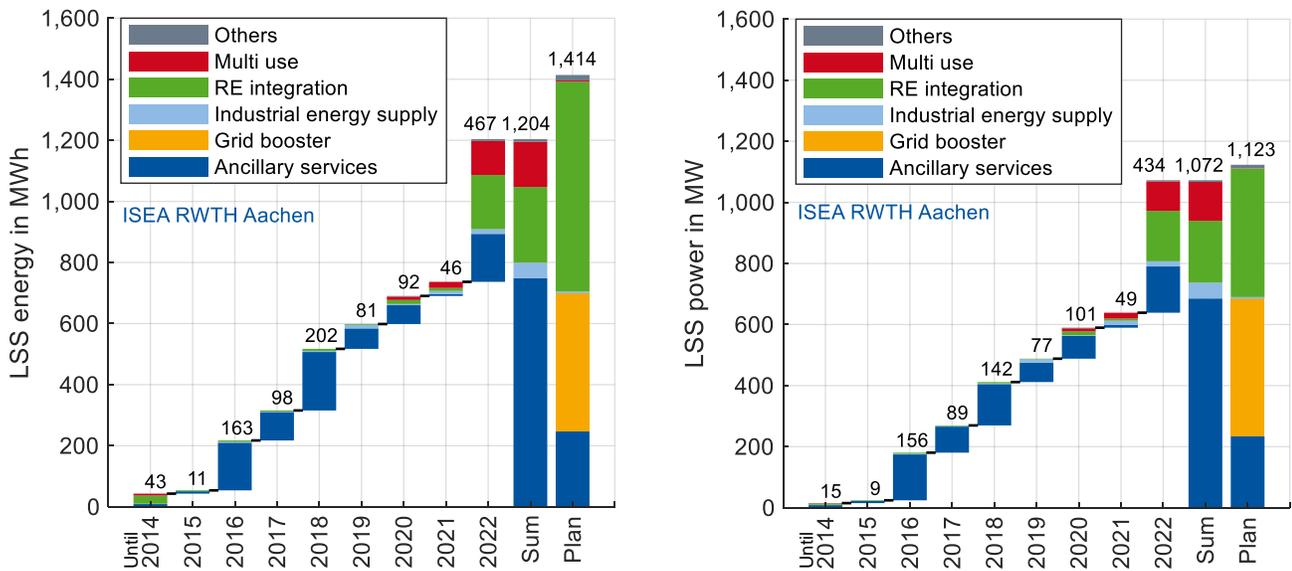

*Figure 23. Development of LSS battery energy (left) and inverter power (right) by use case in Germany, based on own analyses of MASTR DB [34], LSS DB [35], and further research.*

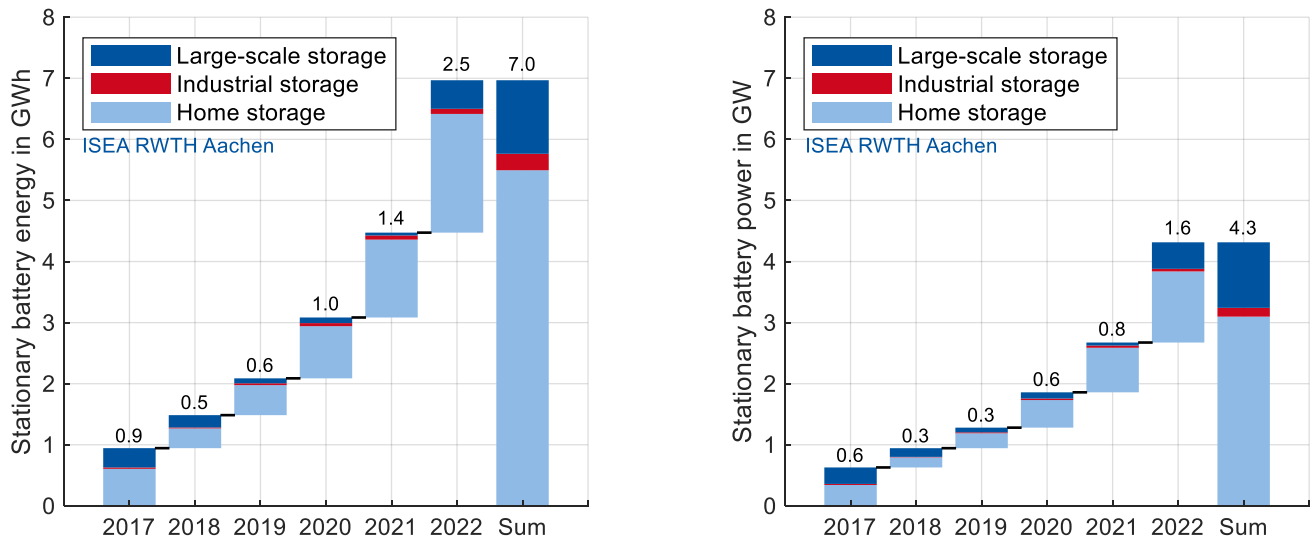

*Figure 24. Development of the stationary battery storage market in Germany based on own analyses of MASTR DB [34], LSS DB [35], and further research. Battery energy (left) and inverter power (right).*

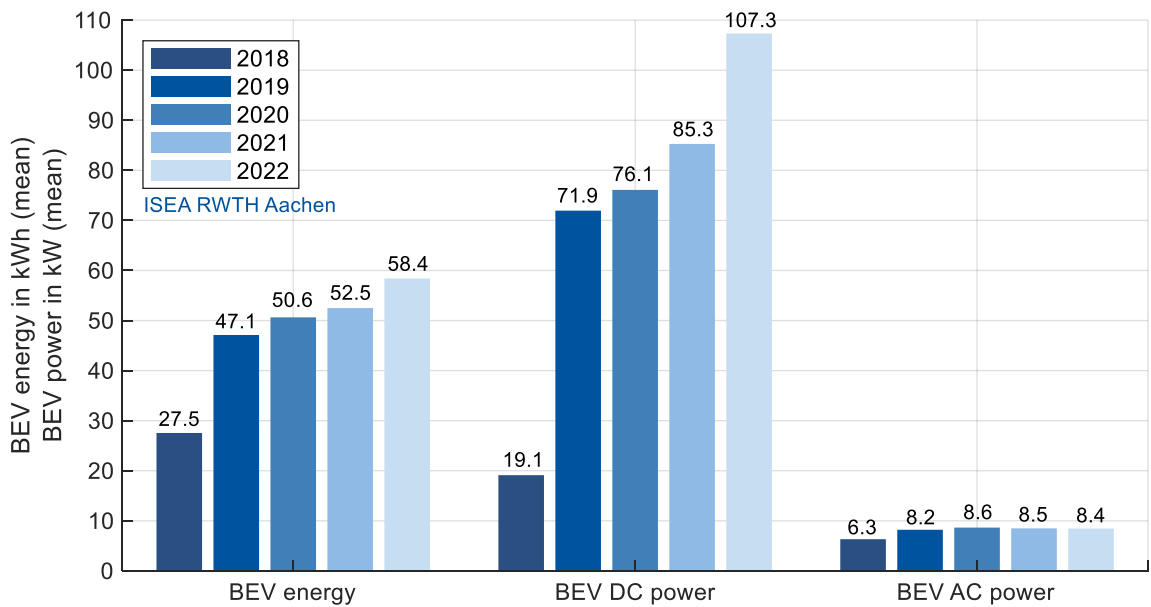

*Figure 25. Mean BEV battery energy and charging power in Germany based on own analyses of FMTA DB [12] and ADAC DB [37].*

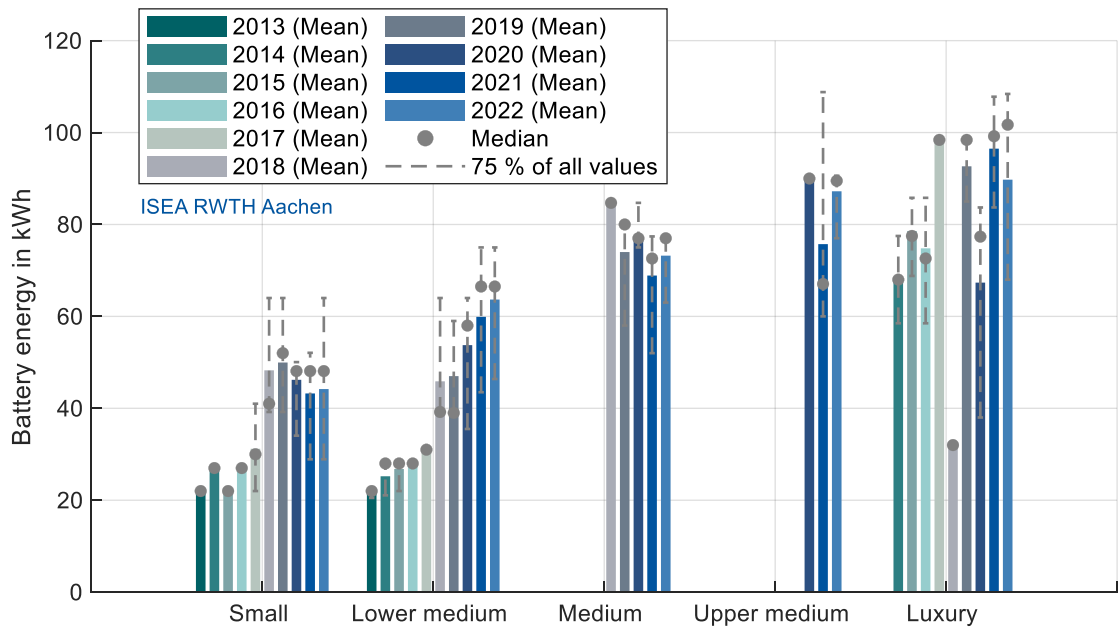

*Figure 26. Battery energy of BEV according to vehicle class based on own analyses of ADAC DB [37]. Values are not weighted by sales volume and show the BEV models that started in the specified year.*

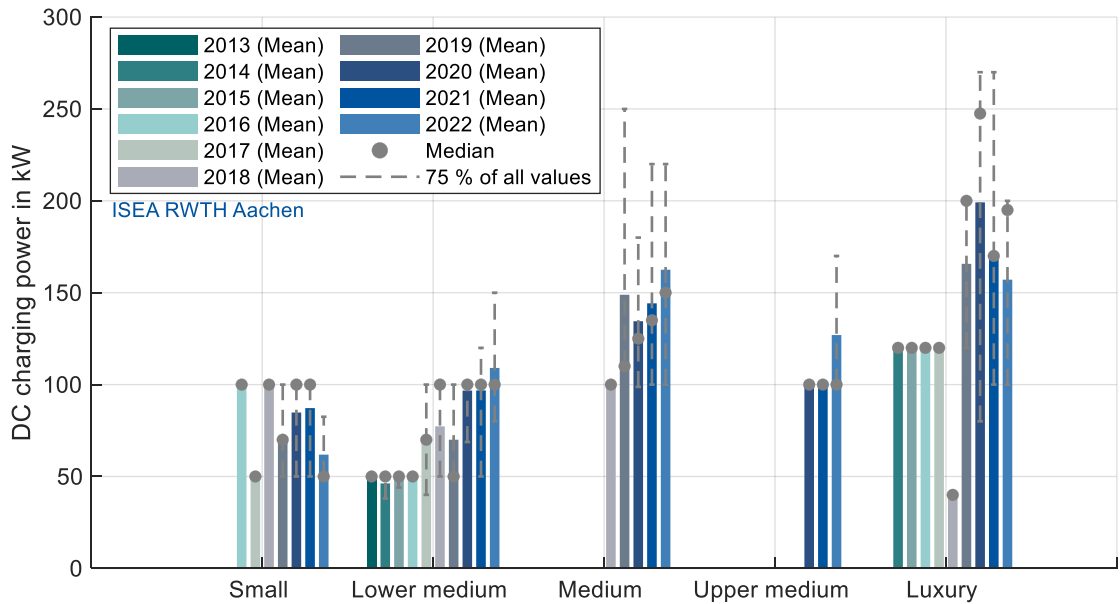

*Figure 27. DC charging power of BEV according to vehicle class based on own analyses of ADAC DB [37]. Values are not weighted by sales volume and show the BEV models that started in the specified year.*

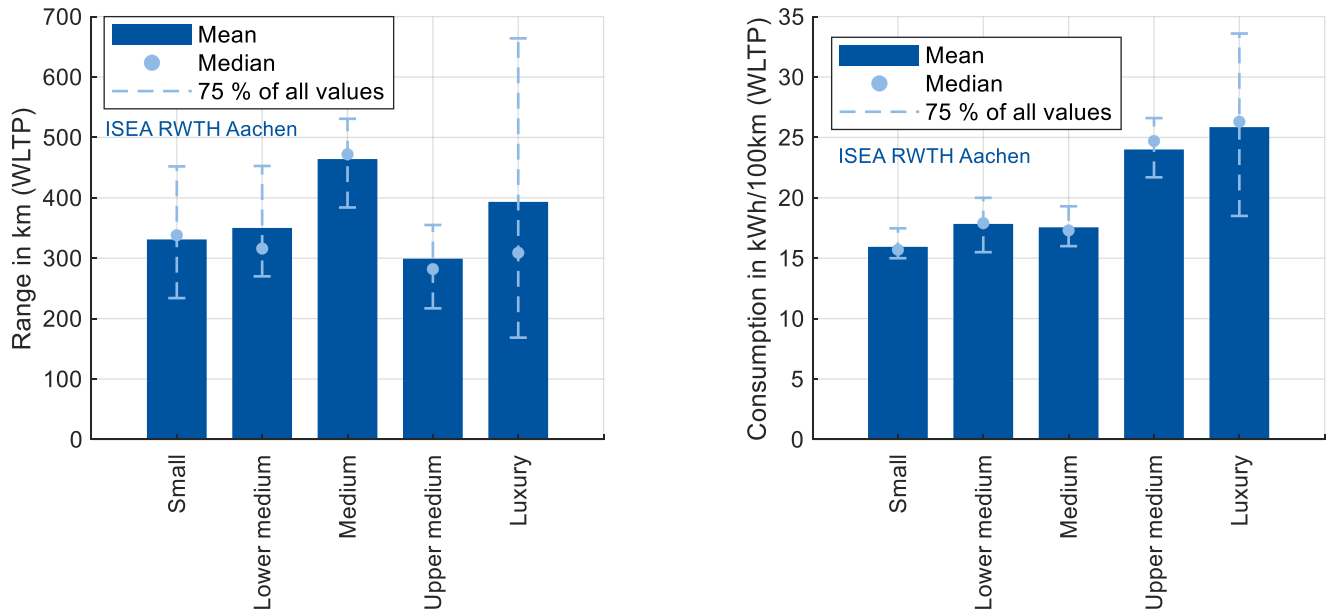

*Figure 28. Range (left) and consumption of new BEV based on own analyses of ADAC DB [37].*
*Values are not weighted by sales volume.*

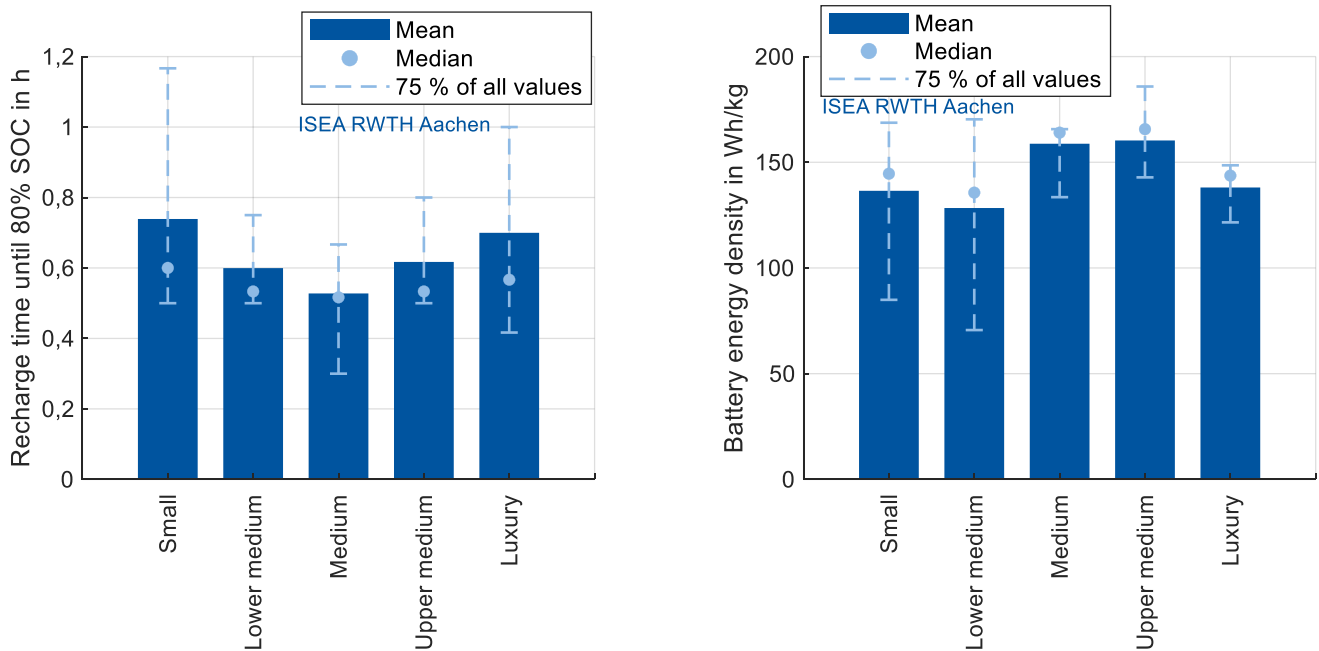

*Figure 29. Recharge time from 0% to 80% with DC charging (left) and energy density (right) of BEV based on own analyses of ADAC DB [37].*
*Values are not weighted by sales volume.*

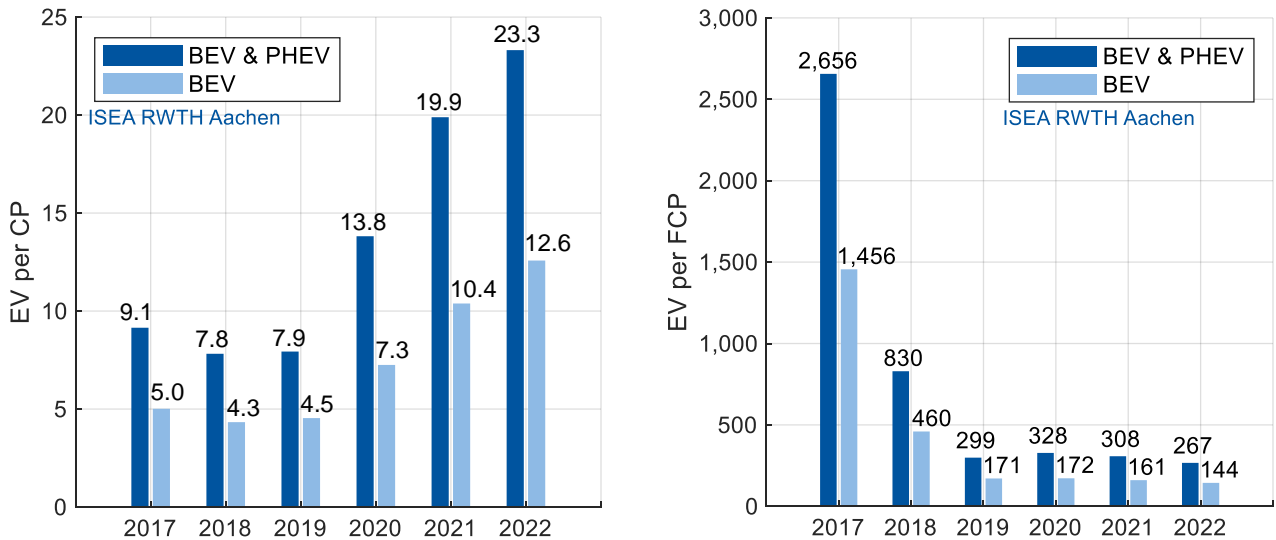

*Figure 30. Number of EV per charging point (CP) (left) and number of EV per fast charging point (FCP) with a power from 150 kW upwards (right) in Germany based on own analyses of FMTA DB [12] and CHARGE DB [38].*

*Table 4: Vehicle classes and examples.*

| Vehicle class | EV examples |
|---|---|
| **Small** | **VW**: Polo, **Opel**: Corsa, **Hyundai**: Kona, **Renault**: Zoe, **KIA**: Soul, **Peugeot**: 2008, **Mini**: Cooper, **Honda**: e |
| **Lower medium** | **Renault**: Kangoo / Mégane, **Fiat**: Doblò, **Opel**: Combo, **Volvo**: (X)C40, **Citroën**: Berlingo / C4, **Nissan**: Leaf / Tonstar, **VW**: Golf / ID.3, **Mercedes-Benz**: Electrive Drive / B250 / EQA, **BMW**: iX1 xDrive, **Ford**: Focus, **Toyota**: Proace City, **KIA**: Niro, **Peugeot**: Partner / Rifter, **Mazda**: MX, **Lexus** UX 300, **MG**: MG 4 / 5 / ZS |
| **Medium** | **Audi**: Q4 e-tron, **Skoda**: Enyaq, **Nissan**: Ariya, **VW**: ID.4/5, **Mercedes-Benz**: EQB/C, **BMW**: i4, iX3, **Ford**: Mustang Mach-E, **Hyundai**: IONIQ 5/6, **KIA**: EV6, **Tesla**: Model 3/Y, **Polestar**: 2, **MG**: Marvel, **Toyota**: bZ4X |
| **Upper medium** | **Mercedes-Benz**: EQE / EQV, **Fiat**: Scudo / Ulysse, **Fisker**: Ocean Hyper Range, **Opel**: Vivaro / Zafira, **Citroën**: Jumpy / Spacetourer, **VW**: ID.Buzz, **BMW**: iX xDrive, **Toyota**: Proace, **Peugeot**: Expert / Traveller |
| **Luxury** | **Lucid**: Air Dual Motor, **Opel**: Movano, **VW**: Crafter, **Mercedes-Benz**: EQS / Sprinter, **BMW**: i7 xDrive, **Ford**: Transit, **Renault**: Master, **Fiat**: Ducato, **Peugeot**: Boxer, **Citroën**: Jumper, **Tesla**: Model S / X, **Porsche**: Taycan, **NIO**: ET7 |